\documentclass[12pt]{article}

\usepackage[utf8]{inputenc}
\usepackage[T1]{fontenc}
\usepackage[english]{babel}
\addto{\captionsenglish}{%
  
}

\usepackage[left=2.5cm, right=2.5cm, top=2.5cm, bottom=2.5cm]{geometry}

\usepackage{amsmath}
\usepackage{physics}

\usepackage{siunitx}
\usepackage{chemformula}
\usepackage[version=3]{mhchem}

\usepackage[super,sort&compress,comma]{natbib}
\usepackage{subfig}

\usepackage{graphicx}
\usepackage{float}
\usepackage{caption}
\usepackage{subcaption}

\usepackage{booktabs}
\usepackage{array}

\usepackage[dvipsnames]{xcolor}
\usepackage{hyperref}
\hypersetup{
    colorlinks=true,
    linkcolor=blue,
    citecolor=blue,
    urlcolor=blue
}

\usepackage{setspace}
\usepackage{comment}
\usepackage{verbatim}
\usepackage{lastpage}
\usepackage{epstopdf}

\newcommand{\todo}[1]{\textcolor{red}{[XXX #1 XXX]}}

\begin{document}

\title{\textbf{Mesoscopic Modeling of Dynamic Tetra-PEG Hydrogel Networks}}

\author{Pietro Miotti$^{1,\ast}$ \and Lucien Cousin$^{2}$ \and Mark W. Tibbitt$^{2}$ \and Igor V. Pivkin$^{1}$}

\date{%
  $^{1}$Institute of Computing, Faculty of Informatics, USI Lugano, Switzerland\\
  $^{2}$Department of Mechanical and Process Engineering, ETH Zürich, Zürich, Switzerland\\[0.5em]
  $^\ast$Correspondence: \href{mailto:pietro.miotti@usi.ch}{pietro.miotti@usi.ch}
}

\maketitle

\begin{abstract}
We introduce a mesoscopic model of dynamic Tetra-PEG hydrogel networks based on a hybrid Dissipative Particle Dynamics/Monte Carlo (DPD/MC) approach. Polymer chains are described by Finite Extensible Nonlinear Elastic (FENE) potential, while reversible cross-links are modeled with Morse potential and Monte Carlo bond exchange governed by Bell’s force-dependent kinetics. After systematic calibration against theory and experiments, the model reproduces the characteristic Maxwell-like viscoelastic response of these networks. In particular, the relaxation time follows the expected scaling, $\tau_R \propto \tau_b (p - p_{\text{gel}})$, and the simulated storage moduli agree with experimental rheology. The mesoscopic resolution allows for graph-based topological analysis, where Tetra-PEG molecules and cross-links are represented as nodes and edges, providing access to bond distributions, fraction of dangling chains, and size of percolating clusters that are challenging to measure experimentally. Comparison with permanent-network predictions further suggests that dynamic bond exchange can affect bond distributions and delay the formation of a system-spanning cluster. This model bridges macromolecular bond kinetics and macroscopic mechanical properties, providing a complementary tool for rational design of dynamic polymer networks. \\
\end{abstract}

\section{Introduction}
Polymer materials play a pervasive role in contemporary society, finding application in packaging, electronics, additive manufacturing, robotics, and biomedical engineering\cite{Jung_Poly_2023}. Traditional polymeric materials are engineered as chemically cross-linked networks, tailored with specific properties for a designated application and lifespan. 
However, this permanent and irreversible nature inherently limits their scope: they exhibit limited reprocessing, self-healing, or adaptation to changing environmental conditions, restricting their possible applications. 
An effective strategy towards the development of a more versatile material not confined to a singular application, involves the synthesis of dynamic polymer networks with reversible bonding interactions~\cite{webber_2022_dynamic}. Within this framework, dynamic covalent chemistry emerges as a promising solution for synthesizing soft materials, leveraging the combination of covalent bond strength and the reversibility of non-covalent interactions~\cite{Rowan_2002_DCN}. The use of dynamic covalent chemistries in polymer networks comprises the class of dynamic covalent networks (DCvNs)  or covalent adaptable networks (CANs)~\cite{Wojtecki_DBA_2010, Kloxin_CAN_2010, Kloxin_CANs_2013, parada_2018_ideal, Winne_DCC_2019, dufort_2020_linking, Friedrichs_DPLS_2024}. In contrast to conventional static networks, these systems undergo continuous bond breaking and reformation. \\ 
Designing DCvNs with specific features and targeted mechanical properties presents fundamental challenges, given that the observed macroscopic viscoelastic behavior emerges as a result of coupled phenomena at play across different length and time scales. This includes effects related to network topology, cross-link density, and the spatial distribution of reversible bonds~\cite{webber_2022_dynamic}. 

Classical analytical frameworks address these relationships through approximations that may not fully capture critical microscopic phenomena and show limited flexibility across different experimental scenarios\cite{Schmid_UMP_2022}. In this context, computational modeling has become a valuable tool to investigate structure--property relationships in complex soft materials\cite{Gautieri2013}.
The choice of the specific computational method depends on the relevant time and length scales of interest as well as the level of desired molecular detail. Continuum-based approaches have proven successful for investigating mechanical properties at macroscopic scales, employing mean-field approximations of microscopic interactions\cite{Lei_RAH_2021}.
Continuum and discrete network models\cite{Hadi_2013_MFN} effectively represent micro--structural features and anisotropic behavior, but operate above the molecular level and cannot fully describe self-assembly or dynamic bond breakage/reformation processes that govern emergent mechanical properties in DCvNs\cite{parada_2018_ideal}.
Particle-based computational models bridge this gap by explicitly representing individual polymer chains and cross-links\cite{Gartner_MS_2019, Rovigatti_NMNI_2019}, spanning resolutions from atomistic molecular dynamics (MD)\cite{Allen_2017_CSL} to coarse-grained representations where particles correspond to molecular clusters\cite{groot_1997_dissipative, groot_2004_applications}. 

For DCvNs specifically, accurate modeling of reversible bond kinetics is essential. 
Several computational strategies have been developed to model reversible bonding (see Ref. \cite{Holoman_review_2025} for a review), including hybrid MD/MC schemes with tunable bond kinetics \cite{Hoy_Fredrickson_2009, Mitra_dybond_2023} and reactive-bond models based on equilibrium statistics \cite{Blanco_ebRxMC_2024}. However, these approaches have been used primarily to study structural and bond dynamic properties, and a direct connection to macroscopic viscoelasticity of dissociative polymer networks remains less developed. Mechanical response has been more investigated in associative networks: swap models for vitrimers \cite{Sciortino_RevCross_2017, Ciarella_RevCross_2022} can access mechanics, but are often implemented with topology-preserving bond exchange. This distinguishes them fundamentally from dissociative networks, where bond breaking creates a temporary topological defect, preventing stress from propagating until a new bond is re-formed, effectively adding a new timescale to the system. Although these methods and related discrete models \cite{Sugimura_2013_MPPN, Takagi_CSNS_2014, formanek_2021_gel, Palkar_2022_CDP, khare_2024_crosslinker} have provided valuable insights, dissociative networks remain challenging because polymer relaxation and bond dissociation occur on different timescales. 

To address this, we decouple structure formation from mechanical probing. We first use polymer dynamics to generate a realistic, percolated network topology. We then apply oscillatory deformation while treating the Tetra-PEG units as rigid bodies and suppressing internal chain modes, so that relaxation is governed by cross-link exchange, consistent with the timescale-separation framework of Parada and Zhao~\cite{parada_2018_ideal}. In this work, we develop a computational model of dynamic covalent networks using Dissipative Particle Dynamics (DPD)\cite{Hoogerbrugge_SMHD_1992}. DPD is a versatile, multi-scale simulation technique for studying coarse-grained biophysical and polymeric systems\cite{symeonidis_2005_dissipative,Zhu_2016_URDPD,Guo_FIT_2011, sirk_EEP_2012}. The method ensures accurate hydrodynamic behavior through conservation of mass and momentum via soft repulsive forces and random thermal fluctuations that satisfy the fluctuation--dissipation theorem\cite{Kubo_FDT_1966}. Further, DPD provides a framework for modeling complex fluid--solid interactions through adjustable conservative, dissipative, and random force parameters that can be tuned to represent different material properties and phase behaviors\cite{groot_1997_dissipative, groot_2004_applications, Espaol_HDPD_1995}. Of note, DPD has been particularly effective in simulating polymer self-assembly and hydrogel systems\cite{nikolov_2018_mesoscale,Liu_2021_DPDCR,Lei_2021_HNM,Palkar_2022_CDP,Li_2023_SADR,khare_2024_crosslinker}. A detailed explanation of DPD is reported in Section~\ref{appendix1}.

In this manuscript, we elaborate our methodology and computational coarse-grained model for Tetra-PEG hydrogels linked by dynamic covalent boronate esters. We calibrated the polymer chain behavior as well as the bond and cross-link behavior using experimental data. We then validated the model through oscillatory experiments, capturing the characteristic viscoelastic behavior of dynamic covalent Tetra-PEG hydrogels. Having established the validity of our DPD-based approach, we investigated two key aspects of dynamic covalent networks computationally. First, we examined how the binding and unbinding kinetics affect the mechanical properties of the material by systematically varying specific model parameters. Second, we showed how our model, combined with graph-theory analysis, enables a more thorough investigation of network topology, from cluster formation to topological defects such as loops and dangling bonds. Comparison with predictions for permanent network formation further suggests that reversible bond exchange may influence bond distributions and the growth of the system-spanning cluster relative to permanent gels. Together, these results provide insight into the relationship between molecular-scale dynamics and macroscopic material properties in dynamic covalent networks, improving our ability to engineer materials through macromolecular design.

\\
\section{Methods}

\subsection{Computational Model}
Our hydrogel computational model addresses the inherent multi-scale nature of the system by decomposing it into two distinct modules: the polymer model and the cross-link model. 
The polymer model represents individual polymer arms that are then assembled into Tetra-PEG units, each composed of four arms. This model provides the structural foundation of the network architecture and governs force propagation throughout the material.

The cross-link model captures the dynamic behavior of the network by determining how bonds between Tetra-PEG units break and reform. Ideal reversible networks, including DCvNs, exhibit Maxwell model behavior~\cite{parada_2018_ideal, dufort_2020_linking}, which is characterized by an elastic response to deformation at short time scales ($t < \tau_R$), followed by a gradual viscous response as bonds dynamically break and reform at longer time scales ($t > \tau_R$). The characteristic relaxation time $\tau_R$ governs the transition between solid-like (elastic) and liquid-like (viscous) behavior and is proportional to the average lifetime of cross-link bonds~\cite{Sheridan_SRR_2012}. Therefore, while the polymer model determines force propagation pathways, the cross-link model, and its associated relaxation time, controls the relaxation of the network through bond breakage events and gives rise to the macroscopic Maxwell behavior. We employed both theoretical frameworks and experimental validation to calibrate and verify both model components, ensuring that their individual and combined behavior accurately reproduced experimentally observed material properties.

\subsubsection{Polymer Model} 
\label{polymer_model}
\begin{figure}[h] 
\centering \includegraphics[width=0.6\columnwidth]{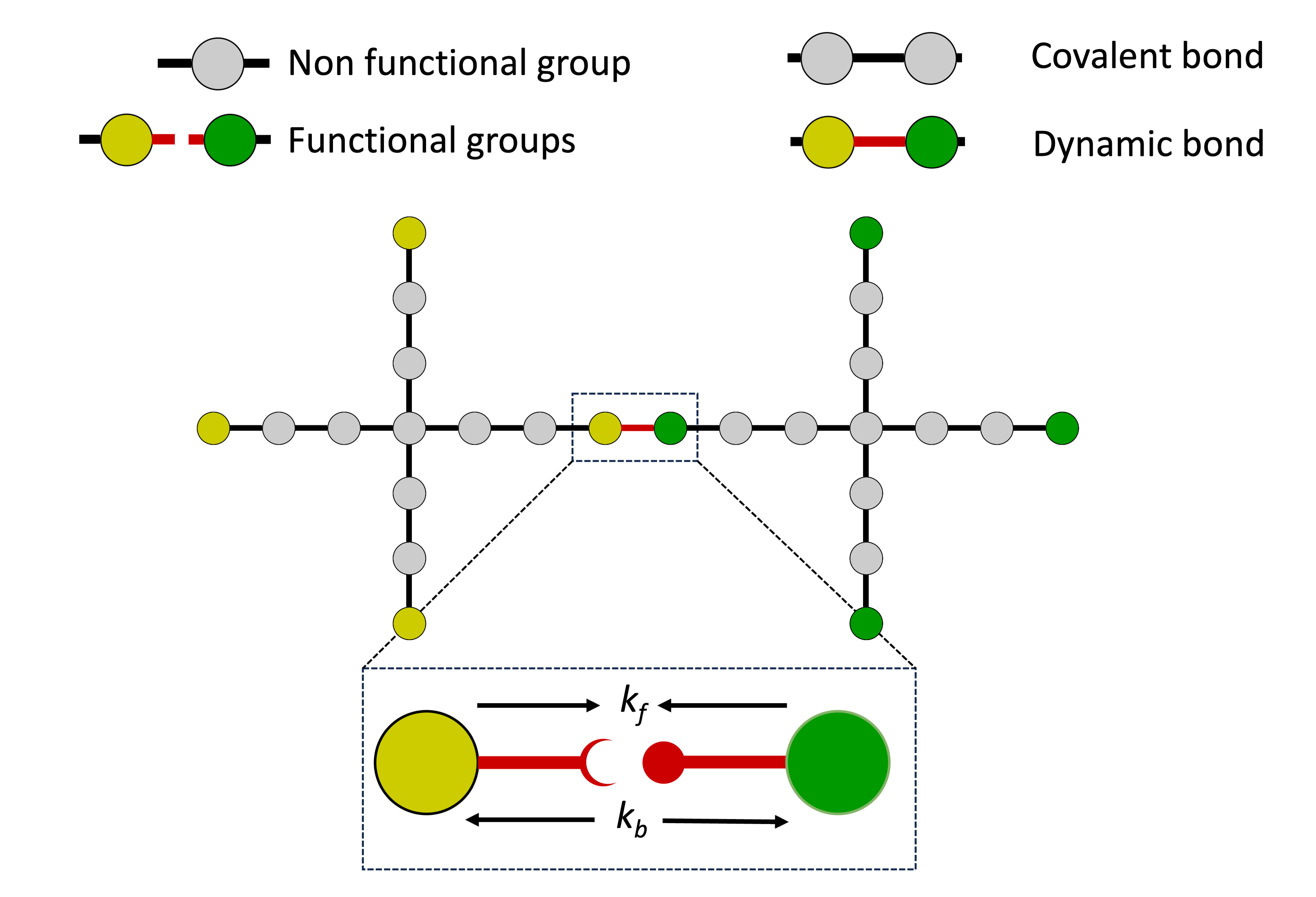} \caption{Schematic representation of the dynamic polymer network model. The figure shows the overall network structure consisting of non-functional (grey) and functional (yellow and green) particles connected by covalent bonds (black) and dynamic cross-links (red). The inset detail illustrates the dynamic bonding mechanism between functional groups, where the association and dissociation rates are governed by forward ($k_f$) and backward ($k_b$) rate constants, with the probability of unbinding depending on the force exerted on the bond.}
\label{fgr:models} 
\end{figure}

We developed a DPD-based polymer model as the basic building block for our Tetra-PEG system. Each Tetra-PEG unit consists of four polymer arms connected in a star shape, as shown in Fig.~\ref{fgr:models}. This model represents the network structure and controls how forces propagate through the material.
Each polymer arm contains DPD particles with two different roles: \textit{functional} and \textit{non-functional} particles. Only \textit{functional} particles can form cross-links with particles from other Tetra-PEG units. To match the selective binding seen in real hydrogels, we use two types of \textit{functional} particles. Cross-links can only form between functional particles of different types. Due to their different roles, functional and non-functional particles were modeled in a different way.

The arms of the Tetra-PEG molecules (Fig.~\ref{fgr:models}) were modeled using the Finite Extensible Nonlinear Elastic (FENE) potential, following the Kremer--Grest bead--spring model,\cite{kremer_1990_dynamics,symeonidis_2005_dissipative} combined with standard DPD forces (namely the conservative, dissipative, and random forces that ensure momentum conservation and temperature control in the system). In this approach, each arm of the Tetra-PEG is represented as a sequence of beads connected by nonlinear springs, with the nonlinear attractive component and the repulsive forces given by the Weeks--Chandler--Andersen (WCA) potential that acts exclusively between connected beads. In this approach, the potential energy for the polymer springs reads as follows:
\begin{equation}
V_\text{FENE}(r) = -\frac{1}{2}KR_0^2\ln\left[1-\left(\frac{r}{R_0}\right)^2\right] + 4\epsilon\left[\left(\frac{\sigma}{r}\right)^{12} - \left(\frac{\sigma}{r}\right)^6\right] + \epsilon
\end{equation}
where $K$ represents the spring constant, $R_0$ is the maximum extension, and $\sigma$ is the characteristic length scale of the beads.

We calibrated the parameters of our coarse-grained polymer model to match the scaling behavior predicted by Flory theory for polymers in good solvents~\cite{flory_1969_statistical,degennes_1979_scaling}. In our coarse-grained representation, particles represent groups of monomers and bonds correspond to effective connectivity between these coarse-grained units rather than individual monomers and covalent bonds. Despite this coarse-graining, the scaling relationship remained valid. The radius of gyration of polymer chains in a good solvent scales with the degree of polymerization $N$ as:
\begin{equation}
R_g \sim N^\nu
\end{equation}
where $\nu = 0.588$ is the Flory scaling exponent~\cite{degennes_1979_scaling}. 
We varied the chain length by constructing linear sequences of $N$ bonded particles. For each chain length, we calculated the corresponding radius of gyration using:
\begin{equation} 
R_g^2 = \frac{1}{N}\sum_{i=1}^{N} (\mathbf{r}_i - \mathbf{r}_\mathrm{cm})^2 
\end{equation}
where $\mathbf{r}_i$ represents the position of particle $i$ and $\mathbf{r}_\mathrm{cm}$ is the center of mass of the polymer chain.

\subsubsection{Cross-link Model}

Modeling dynamic polymer networks requires reversible cross-links between reactive end groups (functional particles of different types) that both propagate mechanical forces and allow topological rearrangement. To achieve this, we developed a hybrid approach combining a Morse potential for continuous force transmission with a Monte Carlo scheme for bond exchange dynamics. Following previous works~\cite{Perego_2022_MDVV,formanek_2021_gel,khare_2024_crosslinker}, the Morse potential describes the mechanical behavior of individual cross-links:

\begin{equation}
V_\text{Morse}(r) = D_0\left[1-e^{-\alpha(r-r_0)}\right]^2
\end{equation}

where $D_0$ is the well depth, $\alpha$ controls the potential width, and $r_0$ is the equilibrium bond distance. This potential provides continuous force transmission through stretched bonds and naturally incorporates bond dissociation through its finite energy barrier. Bond selectivity is ensured by restricting each reactive site to form only one cross-link at a time, preventing multiple simultaneous bonding interactions. Despite these advantages, the potential has an important limitation in modeling dynamic networks: once formed, cross-links remain topologically fixed even when new connections would be favorable, preventing bond exchange. 

To resolve this and enable network reorganization, we implement a Monte Carlo scheme every $N_{mc}$ DPD steps. In this probabilistic approach, bond breaking and formation events are determined by random sampling based on calculated probabilities. This scheme introduces coarse-grained kinetic parameters that directly control association ($k_f$) and dissociation rates ($k_b$), providing flexibility that cannot be achieved by only adjusting temperature or Morse parameters. The frequency of bond updates ($N_{mc}$) affects only the rate at which the system approaches equilibrium but does not influence the final equilibrium proportion of active bonds, which is determined by the balance between formation and breaking probabilities. Bond formation occurs with probability $p_{create} = k^0_f$ between reactive groups within range. Bond breaking follows Bell's model,\cite{bell_1978_models} where mechanical forces reduce the effective dissociation barrier:

\begin{equation}
p_\text{break}(F,r) = \begin{cases}
k^0_b e^{\frac{\lambda F}{k_BT}} & \text{for } r < r^{break}_c \\
1 & \text{for } r \geq r^{break}_c
\end{cases}
\label{eq:probbreak}
\end{equation}

Here, $k^0_b$ is the intrinsic breaking rate, $\lambda$ is the mechanical sensitivity, and the force $F$ derives directly from the Morse potential gradient. 

\begin{figure}[htbp]
    \centering
    \includegraphics[width=0.6\textwidth]{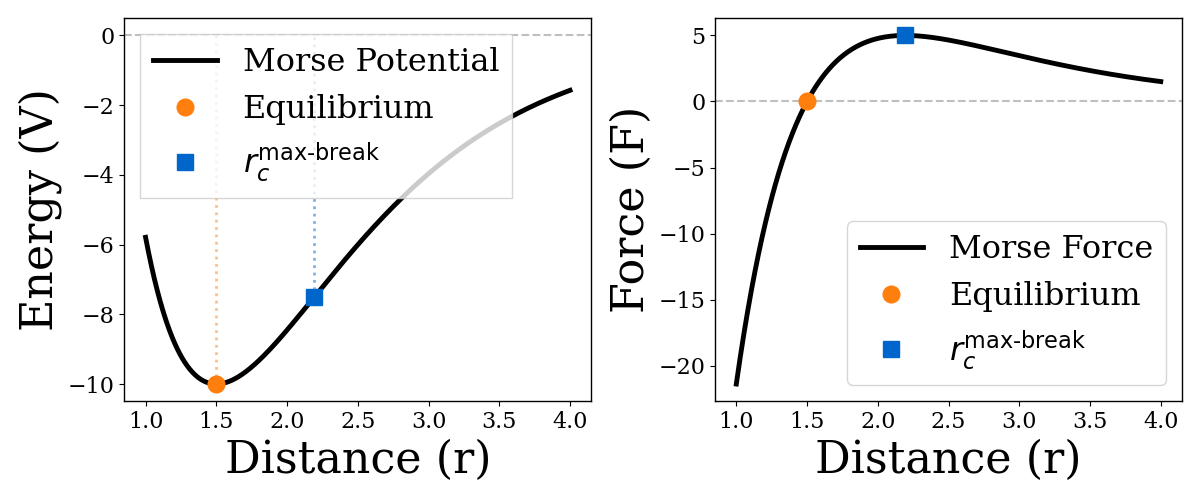}
    \caption{Morse potential and force curves illustrating the natural dissociation criterion. 
    \textbf{Left:} The Morse potential energy as a function of distance, showing the equilibrium position at $r_0$ (orange circle) where the potential reaches its minimum and the critical breaking distance (red square) at the inflection point. 
    \textbf{Right:} The corresponding force curve, where the equilibrium position corresponds to zero force and $r_c^{\text{max-break}}$ marks the maximum attractive force. Beyond this point, the force decreases with increasing separation, indicating loss of binding character and providing a natural cutoff for bond dissociation in molecular dynamics simulations.}
    \label{fig:cutoffrcplot}
\end{figure}

Coupling these two approaches requires defining a physical dissociation limit. The Morse potential's force reaches a maximum before decreasing at larger distances (Fig.~\ref{fig:cutoffrcplot}). When combined with Bell's force-dependent breaking probability, this behavior produces an unphysical result: bonds would become increasingly stable as they stretch beyond the force maximum (\textit{i.e.} they would have lower breaking probabilities at higher distances). To resolve this, we define the maximum breaking distance $r_c^{\text{max-break}}$ at the force maximum, corresponding to the inflection point of the potential:

\begin{equation}
r_c^{\text{max-break}} = r_0 + \frac{\ln(2)}{\alpha}
\end{equation}

This point naturally separates bound states ($r < r_c^{\text{max-break}}$) from dissociating states ($r \geq r_c^{\text{max-break}}$), where bonds are considered broken. We set $r_c^{\text{break}} \approx r_c^{\text{max-break}}$ in our simulations.

\subsubsection{Initialization: Development of the Tetra-PEG Structure.}
\begin{figure}[htbp!]
    \centering
\includegraphics[width=0.6\textwidth]{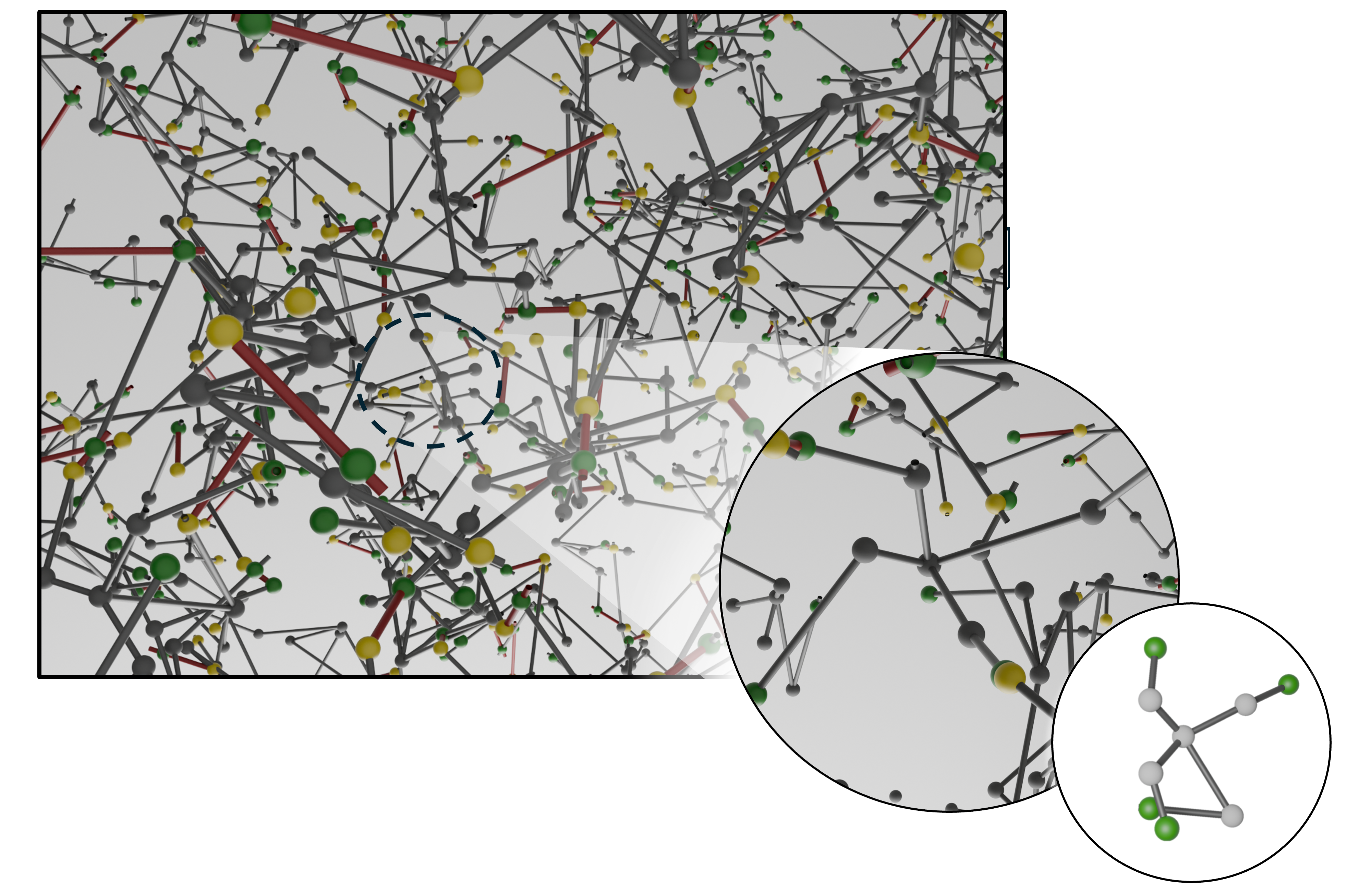}
    \caption{Multi-scale structural characterization of the hydrogel network. The image shows the hierarchical organization from the macroscopic gel structure (left) to the mesoscopic network architecture (center) and microscopic polymer chain arrangement (right).}
    \label{fig:hydrogel_multiscale}
\end{figure}

Using the polymer model and cross-link mechanisms described above, we constructed the dynamic covalent network through DPD simulations of Tetra-PEG polymers. The molecular architecture of a single Tetra-PEG consisted of a central bead with four arms, each arm comprising 2 DPD beads (non-functional) and terminating with reactive end groups (3 beads in total). Each arm behaves like a Gaussian chain, following Flory theory and the parametrization described in the previous section. This coarse-graining level of two bonds per arm was chosen to optimize the molecular size relative to the simulation box, enabling a sufficiently high number of molecules even in moderately sized systems. To ensure our model accurately represents experimental conditions, we adjusted the number of Tetra-PEG in the system to match the experimental volume fraction of the star polymers. We first computed the radius of gyration ($R_g$) of a single Tetra-PEG molecule at equilibrium using the solvent and polymer parameters from the previous section to maintain proper Flory scaling. Using this radius of gyration $R_g$ and the experimental volume fraction of Tetra-PEGs in real solutions ($\rho_\text{exp}$), we then calculated the required number of Tetra-PEG molecules ($N_\text{tetra}$) in the simulation box. The initial configuration was generated using a custom Python script that randomly distributed the Tetra-PEG molecules uniformly throughout the simulation domain, maintaining fixed bond lengths between consecutive beads. We then implemented a two-stage equilibration protocol. The system was first equilibrated for $10^4$ DPD time-steps with no bond formation or breakage dynamics, allowing the Tetra-PEG molecules to reach their equilibrium conformations. Subsequently, dynamic bond creation and breakage mechanisms were activated, and the system was evolved for $5 \times 10^5$ time-steps to establish equilibrium in the number of active (formed) cross-links. This ensured that the network reached a steady-state distribution of active bonds ($p$), forming the hierarchical structure shown in Fig.~\ref{fig:hydrogel_multiscale}.

\subsubsection{Oscillatory Rheometer}
We performed oscillatory shear experiments in order to simulate the mechanical response of our system and to compare it with experimental data. We performed frequency sweeps for frequencies ranging from $\approx 10^0$ rad/s to $\approx 10^{-2}$ rad/s. We used Lees--Edwards boundary conditions, following previous works\cite{Droghetti_DPD_2018,khare_2024_crosslinker} and applied a sinusoidal strain of the form:
\begin{equation}
    \gamma(t) = \gamma_0\sin(\omega t)
\end{equation}
where $\gamma_0$ is the strain amplitude and $\omega$ is the angular frequency. The dynamic moduli were extracted from the stress response through:
\begin{align}
    G'(\omega) &= \frac{\sigma_0}{\gamma_0}\cos(\delta) \label{eq:dynamic_moduli_1} \\
    G''(\omega) &= \frac{\sigma_0}{\gamma_0}\sin(\delta) \label{eq:dynamic_moduli_2}
\end{align}

where $\sigma_0$ is the stress amplitude and $\delta$ is the phase angle between the applied strain and measured stress. This allowed us to directly probe the frequency-dependent viscoelastic response of the network. The stress tensor in the experiment was computed using the virial expression:\cite{thompson_2009_general}
\begin{equation}
    \boldsymbol{\sigma} = \frac{1}{V}\left\langle\sum_i m_i\mathbf{v}_i\mathbf{v}_i + \frac{1}{2}\sum_{i\neq j}\mathbf{r}_{ij}\mathbf{F}_{ij}\right\rangle
\end{equation}
where $V$ is the system volume, $\mathbf{r}_{ij}$ is the relative position vector between particles $i$ and $j$, and $\mathbf{F}_{ij}$ is the force exerted on particle $j$ by particle $i$. 
In our analysis, we focus on the structural contribution to stress arising from active cross-links. Our main interest is the emergence of Maxwell-like behavior in the ideal reversible-network limit, where stress relaxation is governed primarily by the lifetime of reversible cross-links rather than by internal polymer-chain dynamics. According to Parada and Zhao~\cite{parada_2018_ideal}, ideal reversible polymer networks behave like a single Maxwell element: the modulus depends on the equilibrium number of bound cross-links, while the relaxation time is set by cross-link dissociation. Only one main relaxation time is expected because these systems are made of low-molecular-weight macromers near the overlap concentration, where entanglements are negligible and Rouse relaxation is much faster than bond dissociation.

To probe this regime, we use a two-stage protocol. We first equilibrate the system with full polymer dynamics to obtain a percolated network whose conformations reflect equilibrium chain statistics and realistic topology. We then suppress internal polymer relaxation modes during oscillatory shear by constraining each Tetra-PEG molecule as a rigid body, following a strategy previously used in patchy-particle self-assembly studies~\cite{Zhang_TNBB_2003,zhang_2004_self, Padding_CSRS_2009}. In practice, all beads belonging to a given star polymer move together, and the net force on the rigid body is obtained by summing the forces acting on its constituent beads. With internal chain degrees of freedom frozen, the storage modulus reflects the forces transmitted through load-bearing connectivity via active cross-links and vanishes as $p \to 0$. This approach isolates the slow Maxwell relaxation governed by cross-link exchange, rather than the faster response arising from internal polymer motion, which is the main interest of our study.

\section{Results and Discussion}

\subsection{Model Calibration with Experimental Data}
In this section, we describe how the model parameters were calibrated against experimental measurements. We first outline the experimental protocol, then detail the calibration of the polymer model and polymer–solvent interactions, and finally describe the calibration of the cross-link model.

\subsubsection{Experimental Setup}
Aqueous DCvNs were prepared by mixing equimolar amounts of 4-arm PEG stars end-functionalized with either aminophenylboronic acid (PEG-4-APBA) or gluconolactone (PEG-4-GL) in phosphate buffer at 1:1 stoichiometry. Dynamic covalent cross-links were formed through the reversible binding between boronic acid and diol groups.
To systematically vary the proportion of bonds formed ($p$), the concentration of network precursors was varied from 3 to 18~wt\% while maintaining constant stoichiometry and temperature. The chemical equilibrium constant ($K_{\text{eq}}$) for the binding of the boronic acid with gluconolactone was measured using by $^{11}$B nuclear magnetic resonance (NMR) spectroscopy and isothermal titration calorimetry (ITC). From $K_{\text{eq}}$ and the mass concentration of polymer ($c$), the proportion of bonds formed ($p$) at each concentration was calculated using the chemical equilibrium relationship~\cite{parada_2018_ideal}:
\begin{equation}\label{eqn:relation}
p = 1 + \frac{1}{2cK_{\text{eq}}} - \sqrt{\left(1 + \frac{1}{2cK_{\text{eq}}}\right)^2 - 1},
\end{equation} 
This relationship accounts for the thermodynamic behavior of the reversible boronate ester bonds.

\subsubsection{Polymer Model Calibration}
\begin{figure}[t!]
    \centering
    \includegraphics[width=0.6\columnwidth]{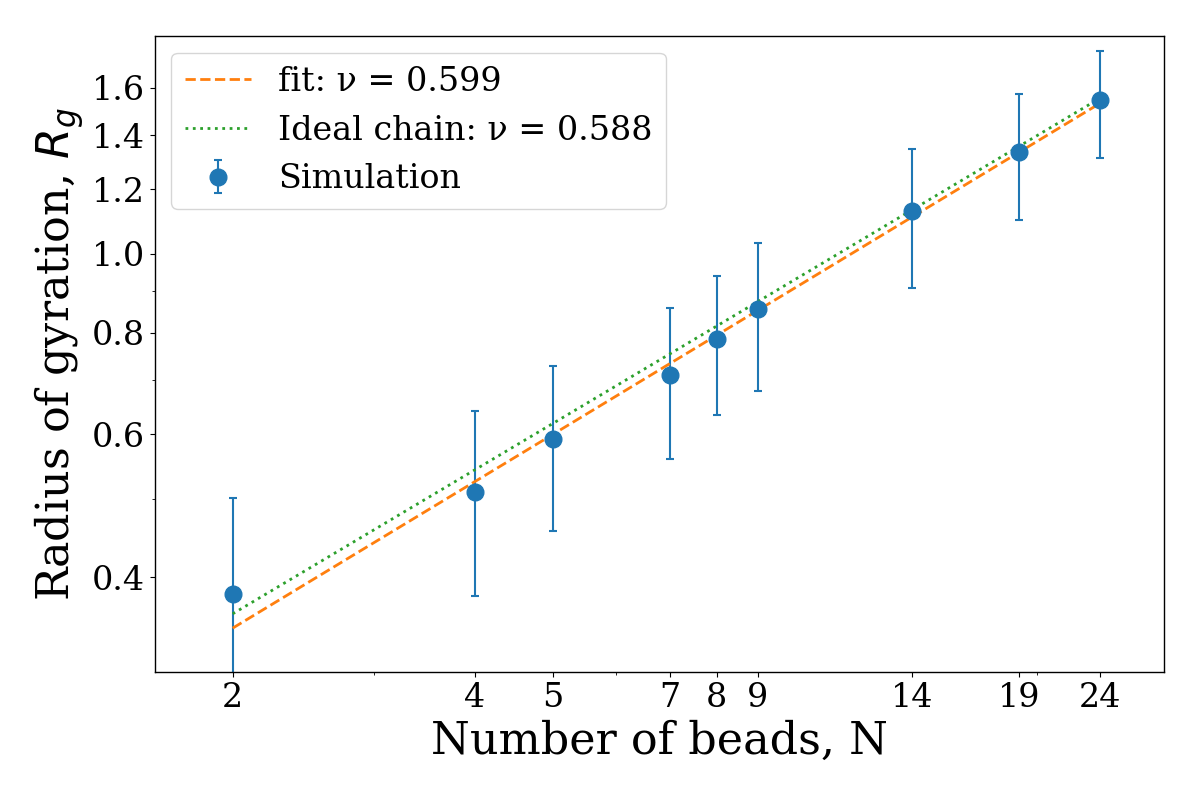}
    \caption{Radius of gyration of Tetra-PEG arms from DPD simulations (orange line) compared with theoretical predictions from Flory's ideal chain model (green line).}
    \label{fgr:chain_stats}
\end{figure}

We first calibrated our coarse-grained polymer model by tuning the polymer and DPD interaction parameters to reproduce the correct polymer scaling behavior in solution. Our system consisted of linear polymer chains immersed in explicit water particles~\cite{Camerin_MRM_2018}, where we systematically adjusted both the conservative force parameters ($a_{\text{ww}}$, $a_{\text{wp}}$, $a_{\text{pp}}$) governing inter-particle interactions and the bond parameters. Our model predicted the expected scaling relationship $R_g \sim N^{0.588}$ for the radius of gyration, which is consistent with a linear chain in a good solvent, \textit{e.g.}, PEG in H$_2$O (Fig.~\ref{fgr:chain_stats}). 

\begin{table}[h]
\centering
\scriptsize
\begin{tabular}{cc}
\hline
Parameter & Value  \\
\hline
DPD-cutoff radius, $r_c$ & 1.00 \\
DPD-viscous force coefficient, $\gamma^{DPD}$ & 4.50\\
repulsion water-water, $a_{ww}$ & 5.00 \\
repulsion water-particle $a_{wp}$ & 2.50 \\
repulsion particle-particle $a_{pp}$ & 5.00 \\
Strength of the bond $K$  & 7.00 \\
Maximum extent of the bond $R_0$ & 2.00 \\
temperature $k_BT$ & 1.00 \\
WCA characteristic length, $\sigma$ & 0.10 \\
WCA depth potential, $\epsilon$ & 1.00 \\
\hline
\end{tabular}
\caption{Parameters for the polymer model}
\label{tab:parameters_polymer}
\end{table}

Complete table of the parameters for the polymer model is reported in Table \ref{tab:parameters_polymer}. For the units of energy, length, mass, and time we follow the DPD parameters, respectively, $k_B T$, $r_c$, $m$, and $t$. In the rest of the article, all numerical values will be given in reduced units where $k_B T = r_c = m = t = 1$. We employed a time-step of $\Delta t = 0.001$.

\subsubsection{Cross-link Model Calibration}
\label{section:cross-link_model_calibration}
\begin{table}[ht!]
\centering
\caption{Experimental characterization of Tetra-PEG networks at different concentrations}
\label{tbl:exp_data}
\scriptsize
\begin{tabular}{cccc}
\hline
PEG concentration & Volume occupied by & Active bonds  \\
(wt\%) & the Tetra-PEGs $\phi$ & (p)  \\
\hline
3  & 0.440 & 0.643 \\
4  & 0.587 & 0.682 \\
5  & 0.734 & 0.710 \\
6  & 0.880 & 0.731 \\
7  & 1.027 & 0.748 \\
8  & 1.174 & 0.762 \\
9  & 1.321 & 0.774 \\
10 & 1.467 & 0.784 \\
11 & 1.614 & 0.793 \\
12 & 1.761 & 0.801 \\
13 & 1.908 & 0.808 \\
14 & 2.054 & 0.814 \\
15 & 2.201 & 0.820 \\
16 & 2.348 & 0.825 \\
17 & 2.494 & 0.830 \\
18 & 2.641 & 0.834 \\
\hline
\end{tabular}
\end{table}

\begin{figure}[h!]
\centering
  \subfloat[]{\includegraphics[width=0.6\columnwidth]{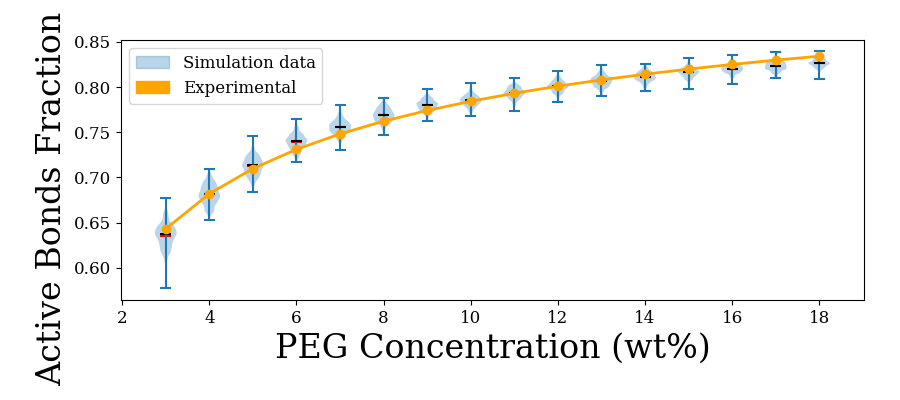}}
  \caption{Comparison between analytical measurements obtained from experimental data and simulation results of active bonds in Tetra-PEG networks. Blue violin plots show the distribution of simulation results obtained using DPD with dynamic cross-linking.} 
  \label{fgr:cross-link}
\end{figure}

With the polymer chain model established, we next calibrated the cross-link model to reproduce experimental bond formation. To achieve quantitative agreement with experiments, we systematically adjusted both the Morse potential parameters and the bond formation/breaking probabilities governing reversible cross-links using experimental data. Our approach involved setting up systems at different polymer concentrations, calculating the expected number of active bonds from the experimental equilibrium constant $K_{eq}$, and then tuning the cross-link model parameters to match these target values across all concentrations. We controlled the system composition by varying the volume fraction of star polymers. Each Tetra-PEG molecule was assembled by connecting four polymer arms at a central junction point. The radius of gyration of each Tetra-PEG molecule ($R_g \approx 0.65$) was calculated from equilibrium simulations using the calibrated parameters from the polymer model described above. We defined the volume fraction as the ratio of the total volume occupied by the spheres delineating the stars to the total volume of box: $\phi = \frac{N_\text{star}V_\text{star}}{V_\text{box}}$ where $V_\text{star}= \frac{4 \pi}{3}R_g^3$. The overlap volume fraction is then $\phi^* = 1$. To determine the target bond fractions for each concentration, we used the chemical equilibrium relationship observed experimentally.
Using a value of $K_{eq} = 841$ $M^{-1}$  measured experimentally, we calculated $p$ from Eq.~\eqref{eqn:relation} for each concentration. The calculated values of $p$ ranged from approximately 0.64 at 3~wt\% to 0.83 at 18~wt\% (Table~\ref{tbl:exp_data}). Fig.~\ref{fgr:cross-link} shows agreement between analytical results derived from experimental measurements and the results from our calibrated computational model across the entire range of polymer concentrations. Each violin plot displays the full distribution of active bonds during the simulation for each concentration, computed after the system reached steady state. The systematic increase in active bonds with polymer concentration is obtained with both experimental and simulation data, showing the characteristic saturation behavior at higher concentrations. 

\begin{table}[h]
\centering
\scriptsize
\begin{tabular}{cc}
\hline
Parameter & Value \\
\hline
Depth of the potential, $D_0$ & 40.00 \\
Equilibrium length $r_{0}$ & 0.10 \\
Width of the potential $\alpha$ & 1.00 \\
Zero-force unbinding rate constant $k_b^0$  & 0.07 \\
binding rate constant $k_f^0$  & 0.9 \\
Mechanical sensitivity for unbinding $\lambda$ & 0.01 \\
MC bond creation/breaking steps $N_{mc}$ & 2500 \\
\hline
\end{tabular}
\caption{Parameters for the cross-link model}
\label{tab:parameters}
\end{table}

\subsection{Validation: Macroscopic Behavior of Dynamic Networks}
We validated our calibrated model by showing that the emergent macroscopic network behavior is in qualitative agreement with experimental observations. Maxwell-like viscoelastic response and scaling relationships developed naturally from the calibrated molecular interactions, demonstrating that our model can be used as a predictive tool for network material design.

\subsubsection{Maxwell-like Viscoelastic Response}

\begin{figure}[htb!]
    \centering
    \begin{minipage}{0.63\columnwidth}
        \centering
        \subfloat[]{\includegraphics[width=1.\textwidth]{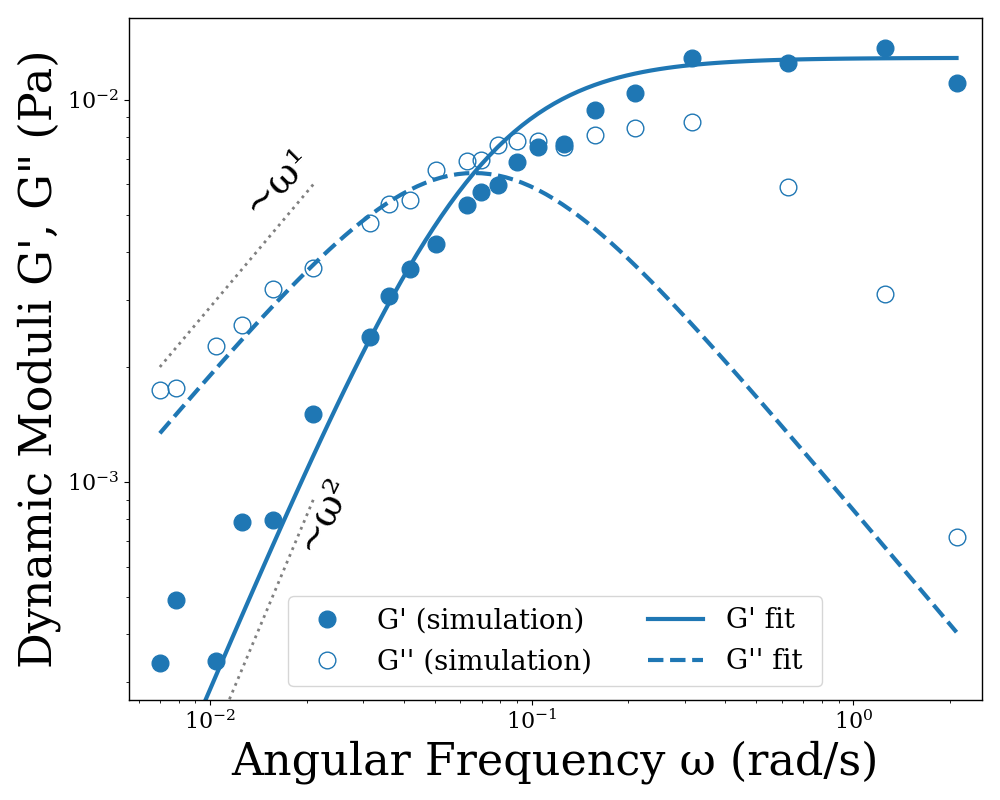}}
    \end{minipage}
    \hfill
    \begin{minipage}{0.36\columnwidth}
        \centering
        \subfloat[]{\includegraphics[width=1.0\textwidth]{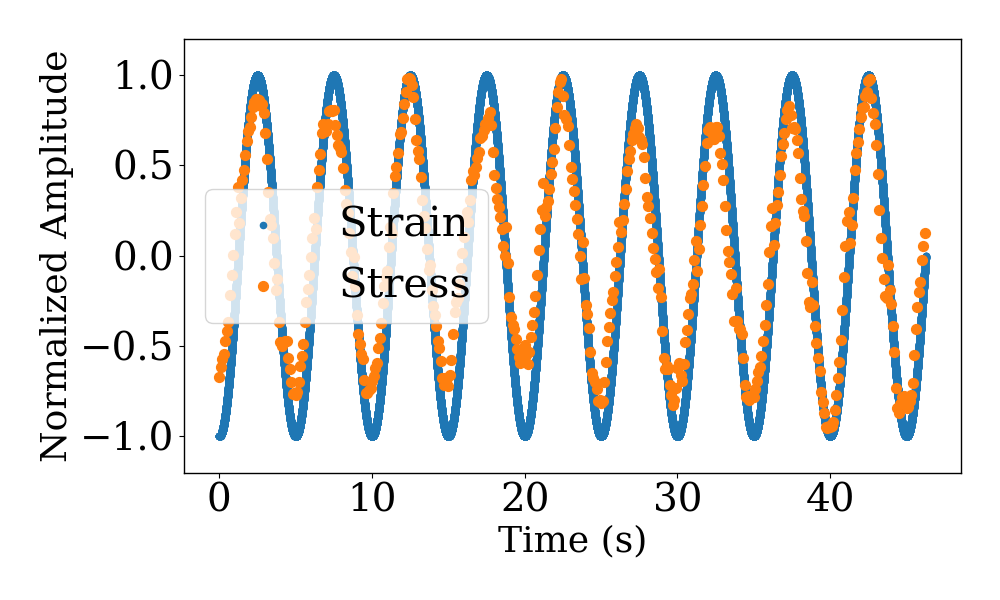}\label{fgr:high_freq}}\\[0.5em]
        \subfloat[]{\includegraphics[width=1.0\textwidth]{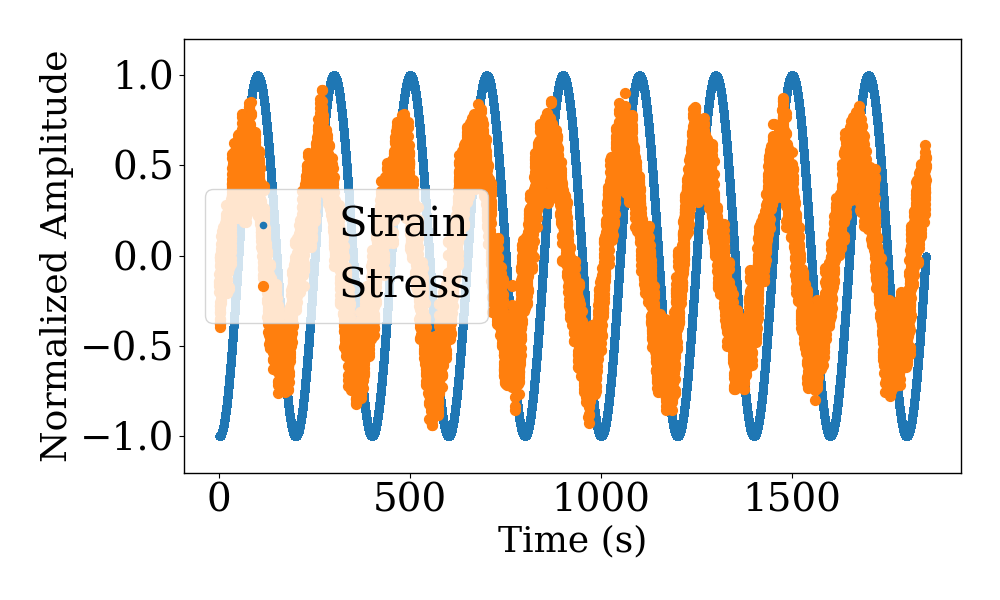}\label{fgr:low_freq}}
    \end{minipage}
    \caption{Maxwell-like viscoelastic behavior of the dynamic hydrogel system. (a) Oscillatory rheology results showing the storage ($G'$) and loss ($G''$) moduli as a function of frequency. The data exhibits Maxwell-like behavior at low frequencies and a plateau at high frequencies. (b) High-frequency oscillatory response ($\omega = 1.256$ rad/s) showing a phase shift between stress and strain. The normalized stress (dashed line) and strain (solid line) signals demonstrate the viscoelastic response of the material. (c) Low-frequency oscillatory response ($\omega = 0.031$ rad/s) demonstrating phase shift approaching $\pi/2$. The normalized stress (dashed line) and strain (solid line) signals illustrate the predominantly viscous behavior.}
    \label{fgr:maxwell_behaviour}
\end{figure}

We simulated oscillatory rheology experiments ($\gamma_0 = 0.38 $; $\omega$ from $\approx 10^0$ rad/s to $\approx 10^{-2}$ rad/s) to investigate the viscoelastic response of our system for a 10 wt\% hydrogel and a simulation box of $20\times10\times8$. To assess finite-size effects, we repeated the analysis in a larger box ($40\times20\times16$) and obtained the same characteristic relaxation time and trends in $G'$ and $G''$. From these simulations, we extracted the storage modulus $G'$ and loss modulus $G''$ across different frequencies. To calculate the moduli, we determined the phase shift ($\delta$) between applied strain and simulated stress response using least-squares fitting of the sinusoidal curves (Fig.~\ref{fgr:maxwell_behaviour}b,c). The temporal delay quantifies the temporal shift between stress and strain oscillations. Using the extracted phase shift and amplitude values, we calculated $G'$ and $G''$ according to Eqs.~\ref{eq:dynamic_moduli_1} and \ref{eq:dynamic_moduli_2}. We repeated this process for all tested frequencies and fitted the low-frequency data points to a Maxwell model using a Python script, determining the plateau modulus and relaxation time as fitting parameters. 
The simulated mechanical response followed Maxwell-like behavior\cite{parada_2018_ideal} in the low-frequency regime, with $G'$ scaling as $\omega^2$ and $G''$ scaling as $\omega$ (Fig.~\ref{fgr:maxwell_behaviour}a). At high frequencies ($\omega = 1.256$ rad/s), the stress response shows a limited phase shift ($\delta$) indicating elastic behavior (Fig.~\ref{fgr:maxwell_behaviour}b), while at low frequencies ($\omega = 0.031$ rad/s), the phase shift approached $\pi/2$, indicating viscous behavior (Fig.~\ref{fgr:maxwell_behaviour}c). Beyond the crossover frequency, the storage modulus $G'$ converged to a plateau value, characteristic of solid-like behavior. The loss modulus $G''$ exhibited a maximum near the crossover frequency, followed by a plateau and subsequent decrease at higher frequencies. The high frequency behavior of $G''$ deviated from the expected Maxwell behavior, as a plateau region is not classically observed. 

The deviation from Maxwell behavior at high frequencies was a consequence of the discrete temporal resolution inherent to the hybrid DPD/MC algorithm. In our approach, the effective bond lifetime is governed by the interplay between physical kinetic parameters---which include the Morse potential parameters ($D_0$, $\alpha$, $r_0$), the mechanical sensitivity ($\lambda$), and the rate constants ($k_b^0$, $k_f^0$)---and the computational Monte Carlo update frequency ($N_{mc}$). While the kinetic parameters represent the physical chemistry of the reversible bonds, $N_{mc}$ imposes a discrete temporal resolution on bond exchanges. At high frequencies, where the oscillation period becomes shorter than the update interval, the bond topology is effectively fixed during the deformation cycle. This results in frequency-independent moduli resembling a covalent network. It follows that the model is specifically intended for timescales $\tau \geq  \tau_R$, where it captures the transition between solid-like and liquid-like behavior as well as the long-term stress relaxation.

In the following sections, we investigate the physical parameters governing network dynamics by systematically varying the rate constants $k_b^0$ and $k_f^0$, which directly control the bond exchange kinetics, while keeping $N_{mc}$ and the other kinetic parameters fixed.

\subsubsection{Characteristic Behaviors of Dynamic Networks}
After ensuring that our system exhibits Maxwell-like behavior, we explored how it reproduces key features of dynamic networks. We first verified that the binding/unbinding process drives the relaxation behavior, as demonstrated in the literature~\cite{Yount_2005_SMD,parada_2018_ideal,dufort_2020_linking}. When we prevented unbinding by setting $k_b^0 = 0$, the system lost its Maxwell-like behavior entirely. Both storage ($G'$) and loss ($G''$) moduli remained nearly constant across all frequencies, with no crossover point (see Section~\ref{Appendix2}). This confirms that binding--unbinding dynamics control the characteristic relaxation time $\tau_R$ of our system, consistent with theoretical predictions~\cite{parada_2018_ideal}. 

\begin{figure}[h!]
\centering
\subfloat[]{%
  \includegraphics[width=0.48\linewidth]{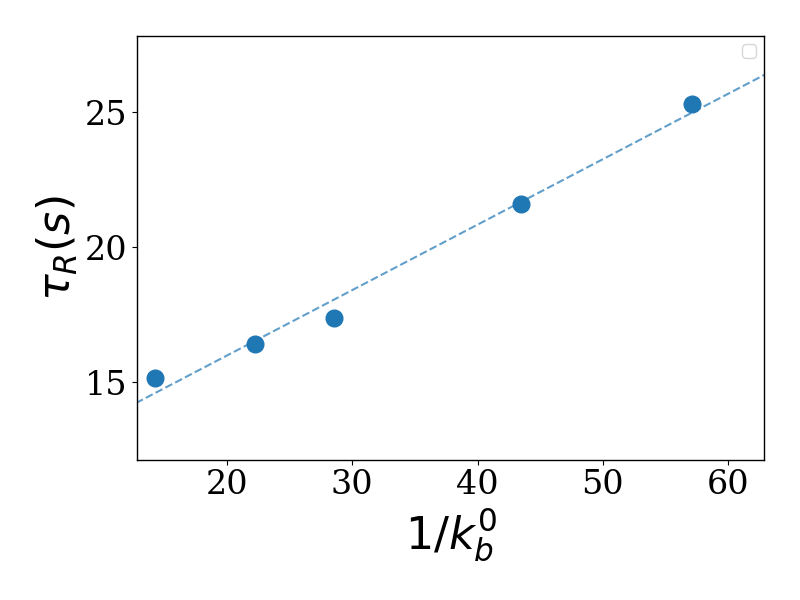}%
  \label{fig:merged_scaling:a}
}\hfill
\subfloat[]{%
  \includegraphics[width=0.48\linewidth]{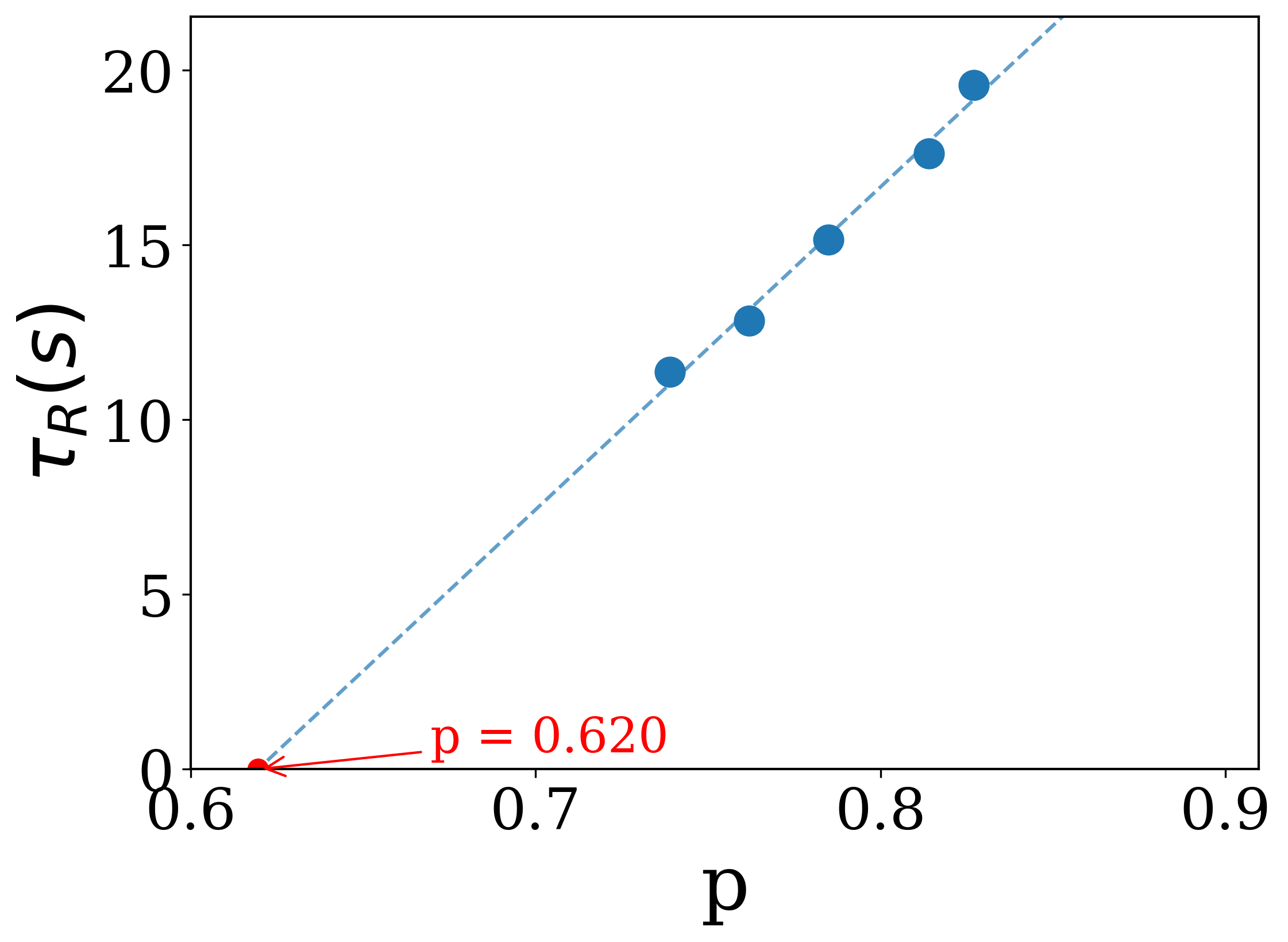}%
  \label{fig:merged_scaling:b}
}\\[0.6em]

\subfloat[]{%
  \includegraphics[width=0.48\linewidth]{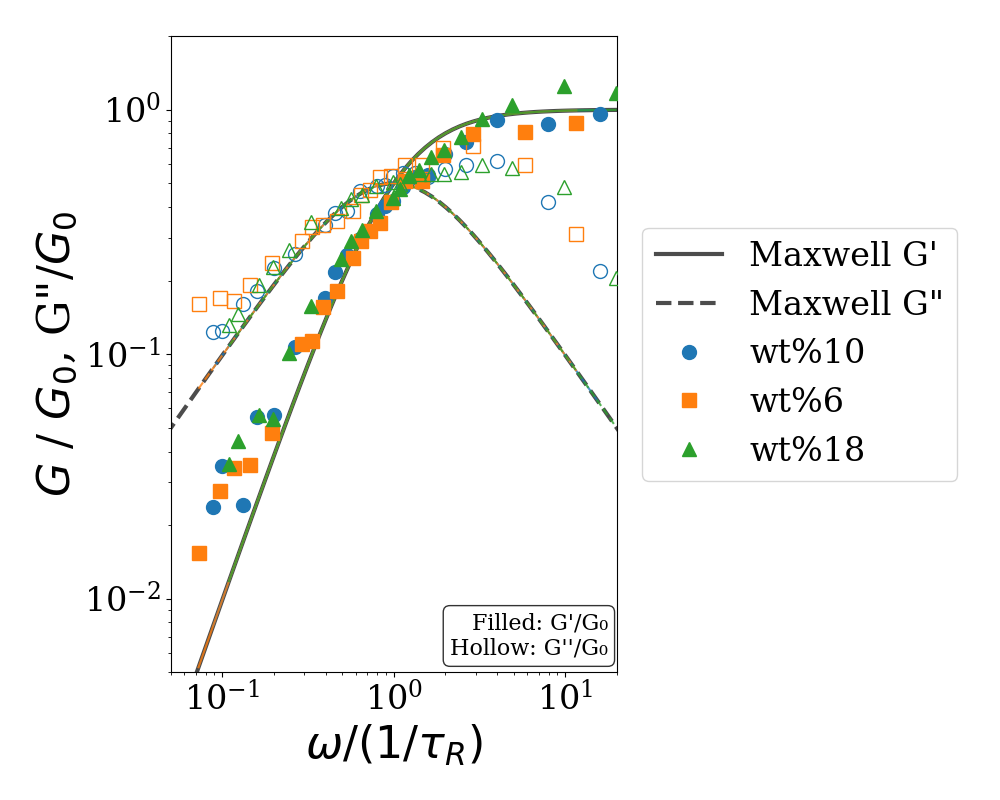}%
  \label{fig:merged_scaling:c}
}\hfill
\subfloat[]{%
  \includegraphics[width=0.48\linewidth]{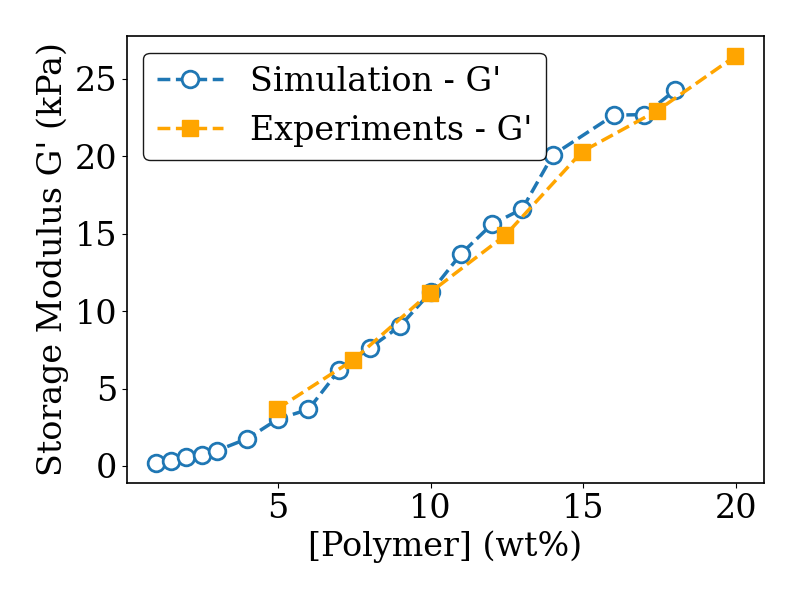}%
  \label{fig:merged_scaling:d}
}
\caption{
Validation of scaling relations in dynamic Tetra-PEG networks.
(a) At fixed active bond fraction \(p\), the crossover frequency scales with \(k_b^0\) (\(\tau_R \propto \tau_b \propto 1/k_b^0\)).
(b) At fixed \(\tau_b\), \(\tau_R\) varies affinely with \(p-p_{\mathrm{gel}}\) (via concentration).
(c) Normalization by \(G_0\) and \(\omega_c\) collapses the viscoelastic response onto a universal master curve.
(d) Storage modulus \(G'\) versus polymer concentration showing agreement between simulations and experiments after normalization at 10 wt\%.
}
\label{fig:merged_scaling}
\end{figure}

With the essential role of bond dynamics established, we investigated how $\tau_R$ depends on the active bond fraction $p$ more systematically. According to theory and experiment, relaxation time in dynamic networks scales as $\tau_R \propto \tau_b (p - p_{\text{gel}})$, where $\tau_b$ is the bond lifetime and $p_{\text{gel}}$ is the percolation threshold~\cite{Semenov_thermo1_1998, Rubinstein_thermo2_1998, Sheridan_SRR_2012, dufort_2020_linking}. To test this relationship, we performed two sets of experiments that independently varied $\tau_b$ and $p$. 
First, we varied the bond lifetime while keeping $p$ fixed. Since $p$ at equilibrium depends on the ratio $k_f^0/k_b^0$ rather than on the absolute values of the rate constants, we systematically changed $k_b^0$ while adjusting $k_f^0$ to maintain a constant ratio $k_f^0/k_b^0 = 12.85$. This approach ensured that the equilibrium bond fraction remained the same for all $k_b^0$, but the time to reach equilibrium varied (\textit{i.e} faster bond exchange dynamics led to more rapid equilibration to $p$, see Section~\ref{Appendix3}). By holding $p$ constant, we isolated the effect of bond kinetics from network connectivity. Under these conditions, the crossover frequency scaled linearly with $k_b^0$ (Fig.~\ref{fig:merged_scaling:a}), consistent with the relation that $\tau_R \propto \tau_b = 1/k_b \propto 1/k_b^0$.

Second, we varied $p$ by changing polymer concentration while keeping the kinetic parameters ($k_f^0$ and $k_b^0$) fixed (Fig.~\ref{fig:merged_scaling:b}).

This modified $p$ without affecting $\tau_b$, and we observed the predicted affine scaling of $\tau_R$ with $p$.  As a consequence of this scaling, when normalized by $G_0$ and $\omega_c$, the viscoelastic responses for different concentrations collapse onto a universal curve (Fig.~\ref{fig:merged_scaling:c}). These results align well with both experimental observations and theoretical predictions for dynamic networks~\cite{dufort_2020_linking, Yount_2005_SMD}.

Finally, we compared simulated storage moduli with experimental data from Marco-Dufort et al.\cite{dufort_2020_linking}. Given the coarse-grained nature of the model, we looked for qualitative agreement rather than absolute numbers; we therefore scaled the simulation data to match the experimental value at 10 wt\%. The resulting comparison, plotted in Fig.~\ref{fig:merged_scaling:d}, shows that the simulated trend aligns well with the experimental data, confirming that our model captures the physics of these dynamic networks.

\subsection{Network Topology Analysis}
Having validated our numerical model by showing that it accurately reproduces experimental results, we then sought to leverage the possibilities offered by numerical models. Recent studies have focused on understanding the effect of molecular-scale features on the macroscopic properties of polymer networks. However, the experimental determination of molecular-scales features, such as loops or clusters, remains extremely challenging~\cite{Lange_conn_2011, Zhou_loops_2012}, and has not yet been achieved in dynamic networks, to the best of our knowledge. Numerical simulations thus provide a powerful tool to investigate and measure the effect of molecular-scale features. Following recent works \cite{vecchio_structural_2021, vecchio_spanning_2022}, we construct a graph representation of the network. By representing each Tetra-PEG molecule as a node and each cross-link as an edge, with edge $(i, j)$ connecting molecules $i$ and $j$, we construct a graph representation of the network. This direct mapping provides molecular-level information about network connectivity, including the distribution of bonds per molecule; topological defects, such as loops and dangling ends; and the formation of network clusters.

\begin{figure}[h!]
    \centering
    \includegraphics[width=0.7\columnwidth]{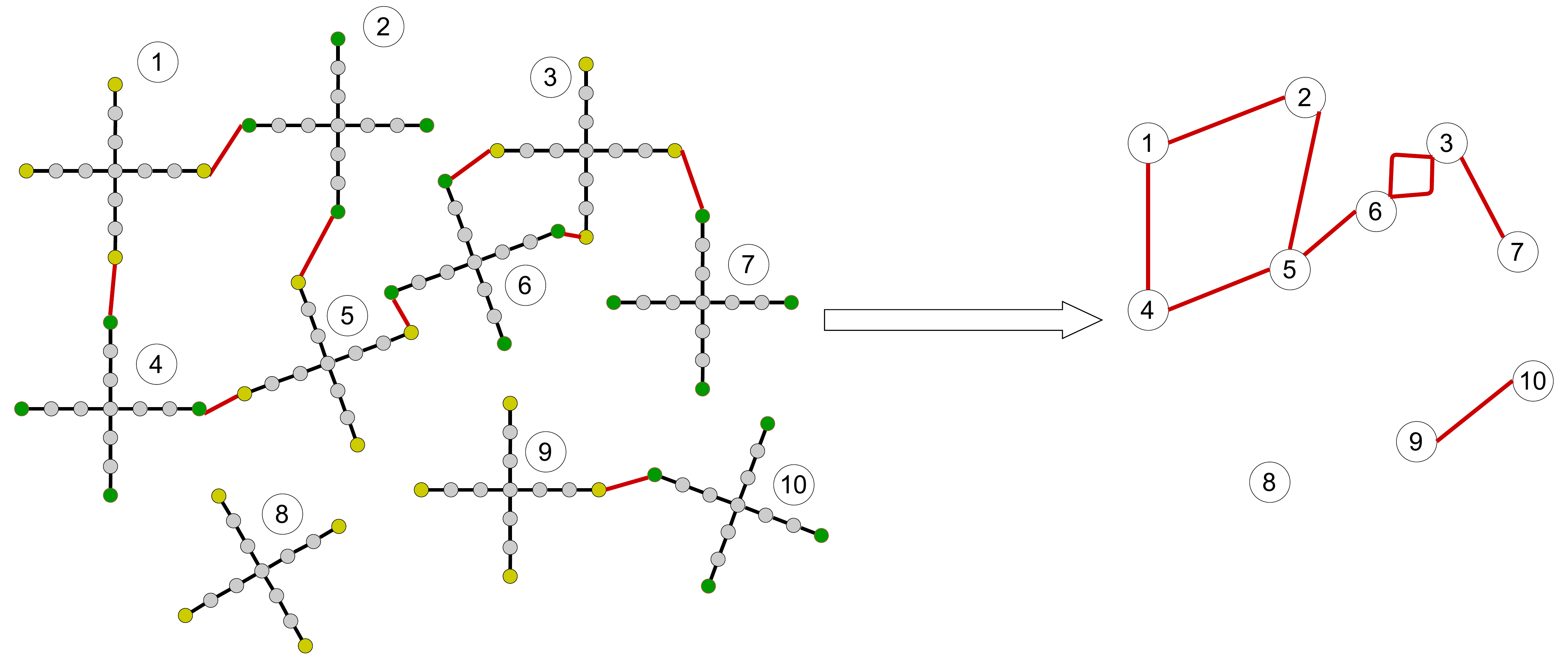}
    \caption{Schematic illustration of the Gel to Graph process. Each Tetra-PEG molecule is represented by a node in the graph located in its center of mass. Each cross-link formed between interacting Tetra-PEG molecules is represented by a weighted edge.}
    \label{fgr:gel_to_graph}
\end{figure}

\subsubsection{Topological defects: Loops and Dangling Bonds}
\begin{figure}[h]
\centering
  \subfloat[]{\includegraphics[width=0.48\columnwidth]{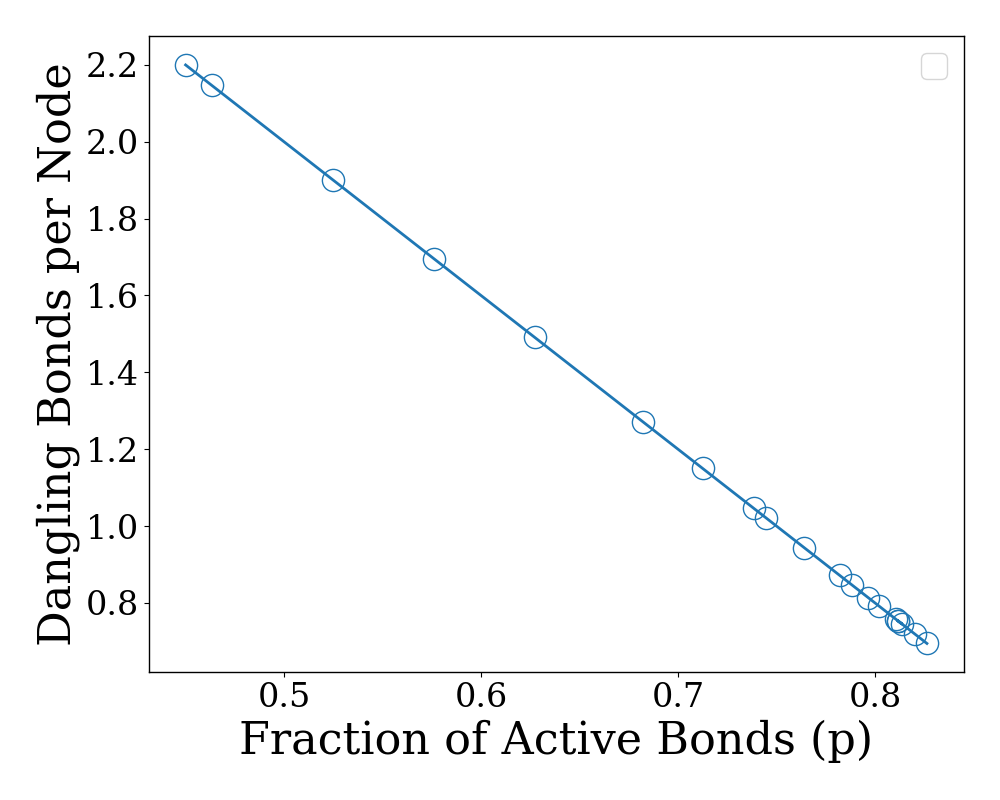}}
  \hfill
  \subfloat[]{\includegraphics[width=0.48\columnwidth]
  {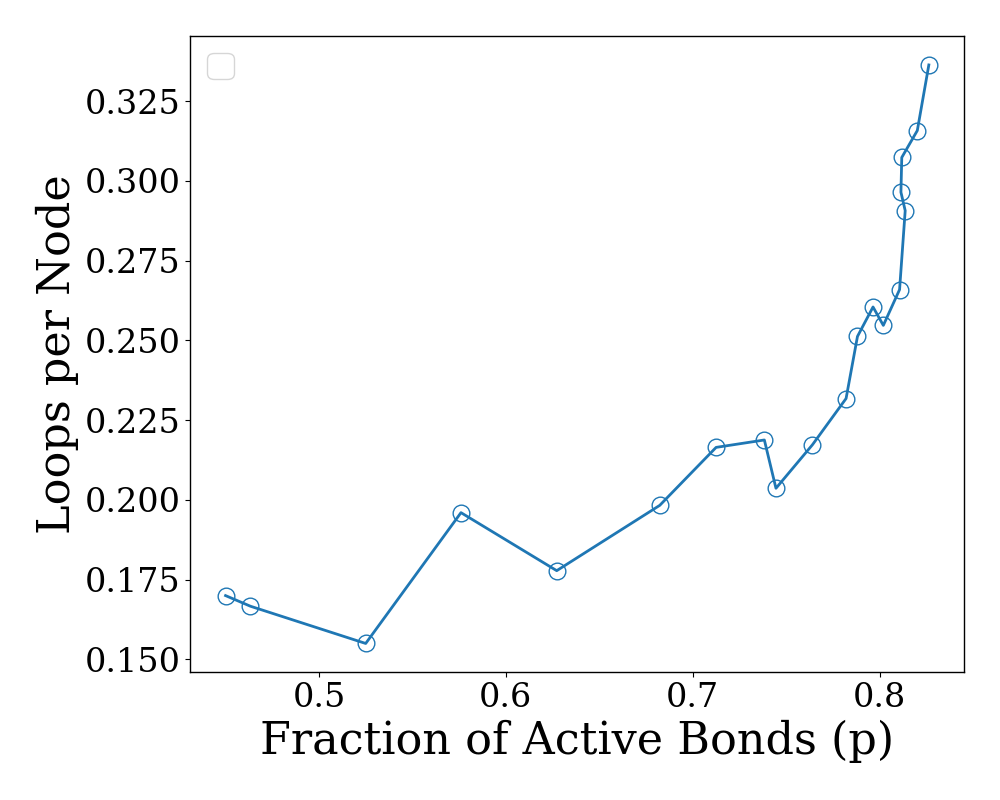}}
  \caption{Topological analysis of Tetra-PEG polymer networks as a function of active bond fraction $p$. (a) Dangling bond density per molecule, showing the average number of unreacted functional groups per PEG molecule as the network evolves. (b) Loop density per molecule }
  \label{fig:dang_loop}
\end{figure}

A first straightforward use of the graph representation is to compute the average number of dangling bonds per Tetra-PEG molecule by subtracting the actual number of bonds from the total possible bonds (where each node has a maximum degree of 4) (Fig.~\ref{fig:dang_loop}(a)). Similarly, second-order loops are found by counting pairs of nodes connected by more than one edge (Fig.~\ref{fig:dang_loop}(b)).

\subsubsection{Bond Distribution}
To characterize the internal structure of the network beyond the average bond fraction $p$, we analyzed the distribution of bonds per Tetra-PEG molecule. We distinguish between two graph representations. In the multigraph, each bond is counted separately, so multiple bonds between the same pair of stars contribute multiple edges. In the simple graph, these multiple edges are merged into a single connection. The multigraph therefore reports how many arms are bonded, whereas the simple graph reports how many distinct stars are connected to a given molecule.

From the simulations, we computed $P_i$ as the fraction of Tetra-PEG molecules with $i$ connected arms, with $i=0,\dots,4$. Equivalently, if $N_i$ denotes the number of molecules with $i$ connected arms and $N$ the total number of molecules, then
\begin{equation}
P_i=\frac{N_i}{N}.
\end{equation}

In order to derive $P_i$, the arms of the same star are classically assumed to react independently of each other \cite{Miller_ndg_1976}. In this case, $P_i$ follows a binomial distribution.

\begin{figure}[ht!]
\centering
\subfloat[]{%
  \includegraphics[width=0.48\linewidth]{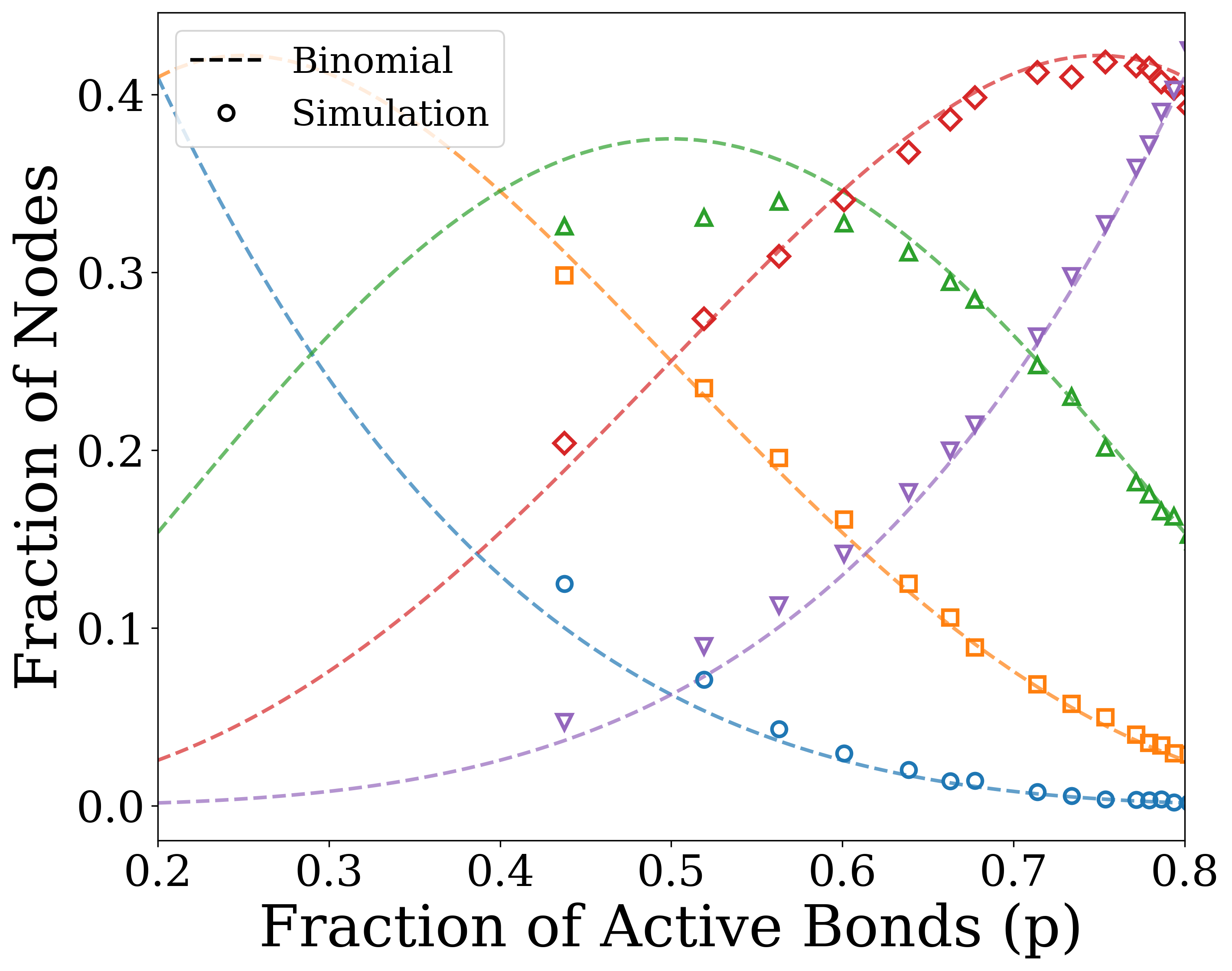}%
  \label{fig:merged_topology:a}
}\hfill
\subfloat[]{%
  \includegraphics[width=0.48\linewidth]{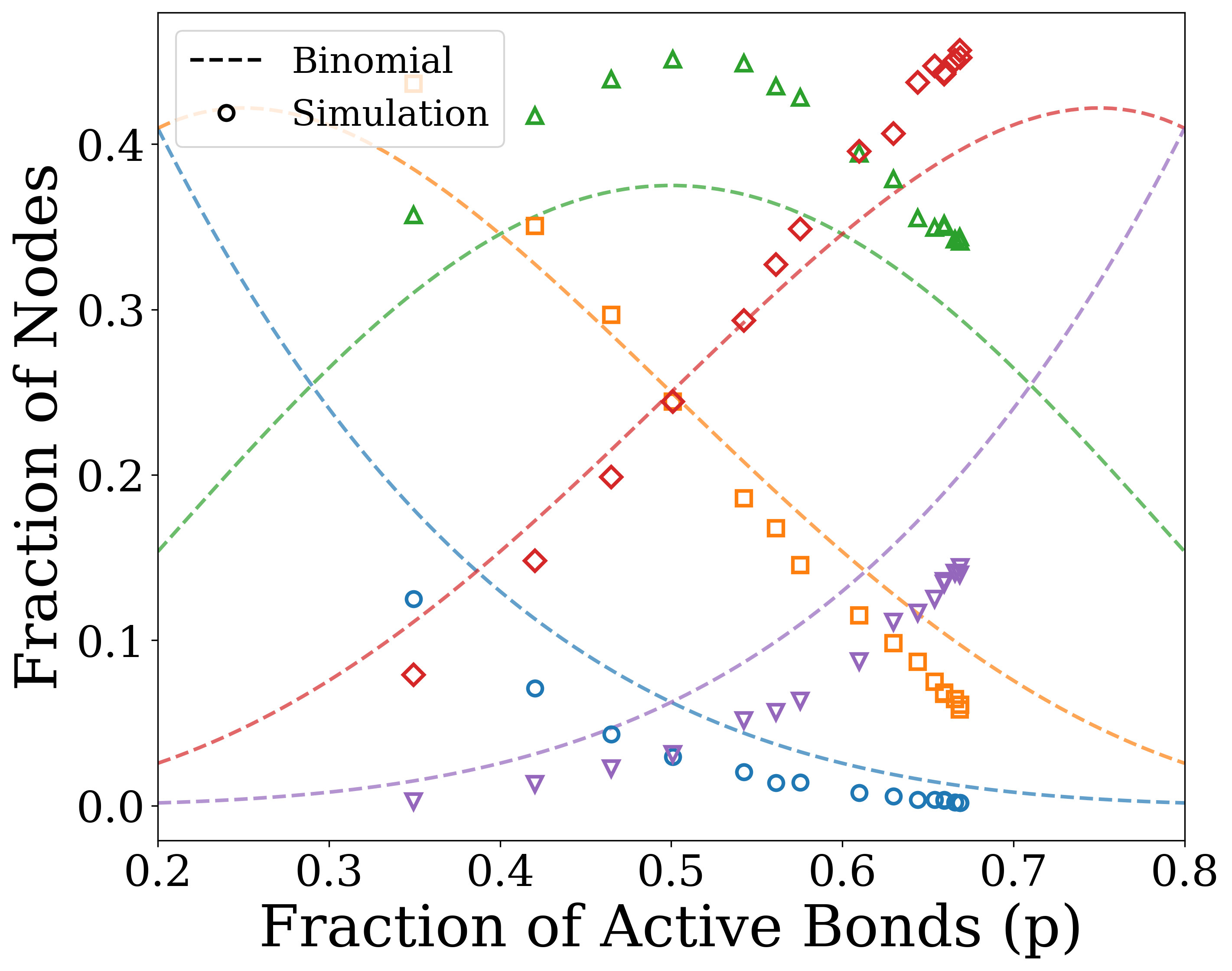}%
  \label{fig:merged_topology:b}
}\\[0.6em]

\subfloat[]{%
  \includegraphics[width=0.48\linewidth]{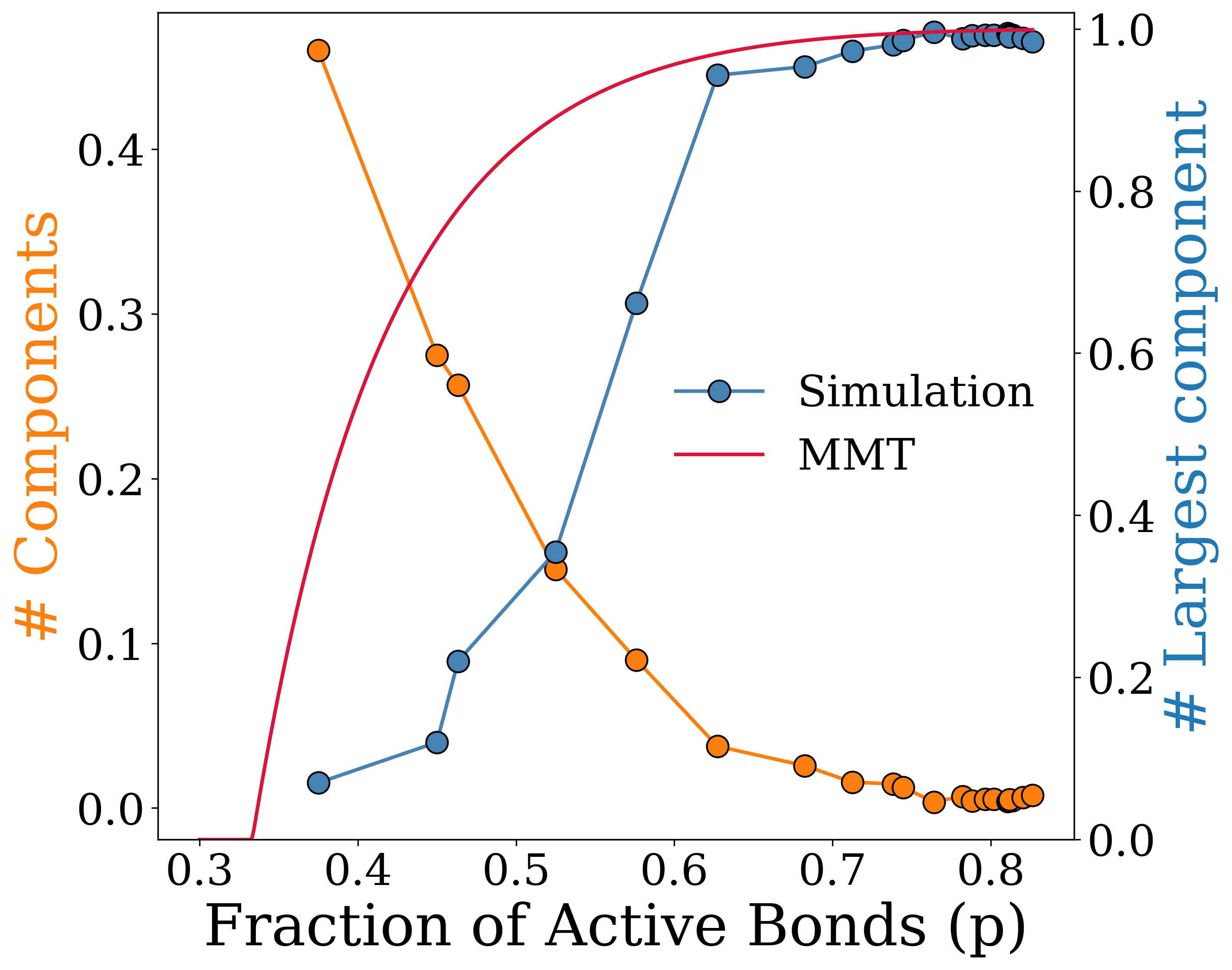}%
  \label{fig:merged_topology:c}
}\hfill
\subfloat[]{%
  \includegraphics[width=0.48\linewidth]{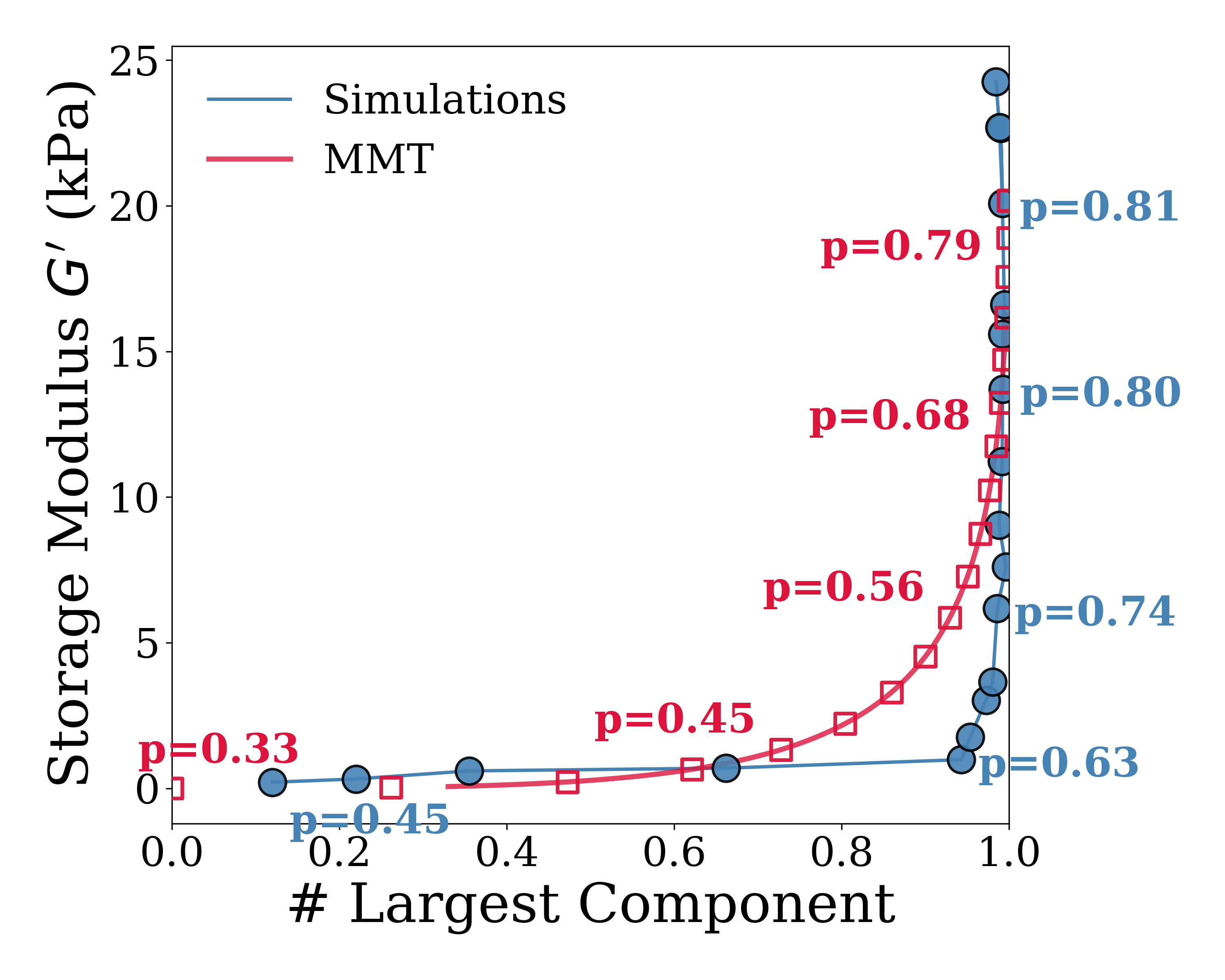}%
  \label{fig:merged_topology:d}
}
\caption{(a) Bond distribution per Tetra-PEG molecule considering loops where colors distinguish bond degree k (blue: k=0, orange: k=1, green: k=2, red: k=3, purple: k=4) (b) Bond distribution per Tetra-PEG molecule discarding loops, considering the simple-graph representation with the same color coding (c) Evolution of the largest connected component fraction and number of disconnected components (see text) as a function of active bond fraction $p$ and comparison with the Miller–Macosko theoretical (MMT) predictions (d) Dependence of storage modulus $G'$ on the fraction of the largest connected component, demonstrating that elasticity emerges only when the network is nearly fully connected (the largest connected component approaches $1.0$). The color scale indicates the active bond fraction $p$.}
\label{fig:merged_topology}
\end{figure}

When measured on the multigraph, the distribution is close to binomial (Fig.~\ref{fig:merged_topology:a}). This indicates that, when all reacted arms are counted, bond occupation is approximately consistent with independent reactivity. However, this representation does not distinguish between bonds that connect a star to different neighbors and repeated bonds between the same pair of stars.

This distinction is important when comparing with Miller--Macosko theory (MMT), which assumes that there are no loops of any size in the network. This situation is better described by using the simple graph, where second-order loops are absent by construction. However, we note that higher-order loops may still be present~\cite{Miller_ndg_1976}. In this representation, the measured distribution deviates from the binomial (Fig.~\ref{fig:merged_topology:b}). Thus, while the number of bonds per star follows a binomial distribution as required by MMT, this appears to no longer hold in a network free of second-order loops. This suggests that the two key assumptions underlying MMT -- (i) independent reactivity of the arms and (ii) a loop-free network -- may not be simultaneously satisfied in dynamic networks. As MMT is widely used to describe the gelation of polymer networks, we next examine the gelation behavior of our systems and compare it with MMT predictions

\subsubsection{Clusters and Gelation}

In order to examine the gelation behavior, we investigated how local connectivity translates into global network structure. Using the graph representation of the simulated gels, we computed both the number of disconnected components and the fraction of molecules belonging to the largest connected component (both normalized by the total number of molecules) as functions of the active bond fraction $p$ (Fig.~\ref{fig:merged_topology:c}). In a simulation, the largest connected component serves as a practical proxy for the percolating (\textit{infinite}) cluster, so its size can be compared with the gel fraction $P_{\mathrm{gel}}$. As expected, increasing $p$ progressively merges small isolated clusters into larger ones. At low $p$, the system is fragmented into many disconnected components, whereas at higher $p$, these clusters coalesce and a dominant connected component emerges.

To compare this behavior with mean-field expectations, we considered the Miller--Macosko prediction for the gel fraction for permanent networks. In that framework, reactive groups are assumed to react independently, and network connectivity is described by a tree-like branching process~\cite{Miller_ndg_1976}. For a tetra-functional network, the gel fraction is given by
\begin{equation}
P_{\mathrm{gel}} = 1 - Q^4,
\end{equation}
where $Q$ satisfies $Q = 1 - p + pQ^3$. Here, $Q$ is the probability that a reacted arm does not connect to the infinite cluster.
The simulated growth of the largest connected component is shifted relative to this prediction (Fig.~\ref{fig:merged_topology:c}), with gelation occurring at a higher bond fraction than expected for an ideal tree-like network. This shift is consistent with the bond-distribution analysis above: although every reacted arm contributes to $p$, bonds that are included in loops increase the total bond fraction without creating new independent pathways. 

To further connect topology and mechanics and predict the onset of gelation, we plotted the storage modulus $G'$ as a function of the fraction of molecules in the largest connected component (Fig.~\ref{fig:merged_topology:d}), comparing it with the MMT predictions.
We observe a steep rise in $G'$ once the network becomes nearly fully connected, which we take as the gel point of our system. Remarkably, the obtained value ($p_{\mathrm{gel}} = 0.63$) agrees well with the gel point extracted from the affine scaling of $\tau_R$ with $p$ ($p_{\mathrm{gel}} = 0.62$ from the x-axis intercept in Fig.~\ref{fig:merged_scaling:b}~\cite{Semenov_thermo1_1998, Sheridan_SRR_2012}), while MMT predicts $p_{\mathrm{gel}} = 0.33$. The evolution of $G'$ as a function of the fraction of stars in the largest connected component also differs from MMT. Taken together, these results indicate that DCvNs behave differently at both the molecular and macroscopic scale than previously assumed, consistent with the work of Cousin et al.~\cite{cousin_entropy_2026} Our simulations thus offer a way to connect microscopic topology to macroscopic properties in DCvNs, and point to new directions for understanding this class of materials.

\section*{Conclusions}

In this work, we developed and validated a computational model for dynamic polymer networks using a combination of theoretical results and experimental data. 
The calibrated model captures the Maxwell-like behavior characteristic of dynamic Tetra-PEG hydrogels and reproduces the expected affine relation $\tau_R \propto \tau_b (p - p_{gel})$. In addition, the elastic modulus shows good agreement with the experimental concentration dependence after normalization. 
We also demonstrated how simulations can provide direct access to structural features that are challenging to measure using current experimental techniques, including topological defects, bond distribution and cluster statistics. Comparison with permanent-network predictions suggests that reversible bond exchange may influence bond distributions and delay gelation. These results support the use of the present framework as a complementary tool for linking network topology, bond dynamics, and macroscopic material response in dynamic polymer networks.

\section*{Author contributions}
P.M. performed the simulations and analysis. L.C. and M.W.T. provided experimental data and validation and contributed to the analysis. I.V.P. supervised the computational work. All authors contributed to writing the manuscript.

\section*{Conflicts of interest}
There are no conflicts to declare.

\section*{Data availability}
The data supporting this article have been included as part of the Supplementary Information.

\section*{Acknowledgements}
I.V.P. acknowledges support from the Swiss National Science Foundation grant 205321\_173020.
Simulations were carried out at the Swiss National Supercomputer Center under projects u4 and lp09. 


\newpage
\appendix
\section{Appendices}
\subsection{DPD Framework}\label{appendix1}
Dissipative Particle Dynamics (DPD) has emerged as a powerful simulation technique for investigating biophysical and polymeric systems~\cite{symeonidis_2005_dissipative, Zhu_2016_URDPD, Guo_FIT_2011}. The method ensures accurate hydrodynamic behavior through conservation of mass and momentum~\cite{groot_1997_dissipative, groot_2004_applications}, while providing a framework for modeling complex interactions between fluid, and solid components through interaction parameters~\cite{Yeh_MMO_2015, Wang_2020_CREN}. The temporal evolution of the system is governed by Newton's equations of motion:

\begin{equation}
\frac{d\mathbf{r}_i}{dt} = \mathbf{v}_i
\end{equation}
\begin{equation}
\frac{d\mathbf{v}_i}{dt} = \mathbf{f}_i
\end{equation}
where $\mathbf{r}_i$ and $\mathbf{v}_i$ denote the position and velocity of particle $i$, respectively. The total force $\mathbf{f}_i$ acting on particle $i$ has three components $f_i = \sum_j F_{ij}^C + F_{ij}^D + F_{ij}^R$:
\begin{enumerate}
\item Conservative force ($\mathbf{F}^C$):
\begin{equation}
\mathbf{F}^C_{ij} = \begin{cases}
a_{ij}(1 - r_{ij}/r_c)\hat{\mathbf{r}}_{ij} & \text{for } r_{ij} < r_c \\
0 & \text{for } r_{ij} \geq r_c
\end{cases}
\end{equation}
where $a_{ij}$ is the maximum repulsion between particles $i$ and $j$, $r_c$ is the cutoff radius, and $\hat{\mathbf{r}}_{ij}$ is the unit vector.

\item Dissipative force ($\mathbf{F}^D$):
\begin{equation}
\mathbf{F}^D_{ij} = -\gamma^{DPD}\omega^D(r_{ij})(\mathbf{v}_{ij}\cdot\hat{\mathbf{r}}_{ij})\hat{\mathbf{r}}_{ij}
\end{equation}
where $\gamma^{DPD}$ is the friction coefficient and $\omega^D$ is a weight function. The weight determines how the conservative, dissipative, and random forces between particles decay with distance.

\item Random force ($\mathbf{F}^R$):
\begin{equation}
\mathbf{F}^R_{ij} = \sigma^{DPD}\omega^R(r_{ij})\xi_{ij}\hat{\mathbf{r}}_{ij}
\end{equation}
where $\sigma^{DPD}$ is the noise amplitude and $\xi_{ij}$ is a random number with zero mean and unit variance.
\end{enumerate}

The weight functions are related by $[\omega^R(r)]^2 = \omega^D(r)$ to satisfy the fluctuation-dissipation theorem. This relationship between the dissipative and random forces ensures proper thermal equilibrium, 
\begin{equation}
\omega^D(r) = \begin{cases}
(1 - r/r_c)^2 & \text{for } r < r_c \\
0 & \text{for } r \geq r_c
\end{cases}
\end{equation}

\subsection{Appendix 2}\label{Appendix2}
\begin{figure}[htbp!]
    \centering
    \includegraphics[width=0.6\columnwidth]{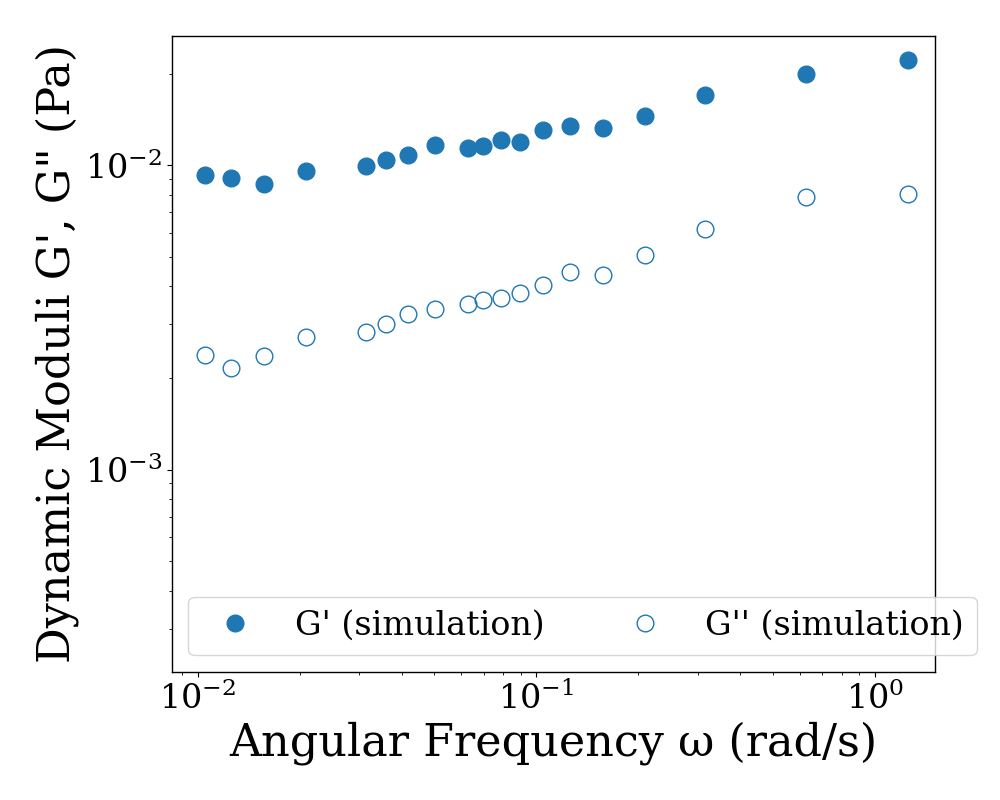}
    \caption{Effect of eliminating unbinding ($k_b^0=0$) on network relaxation. Without unbinding, $G'$ and $G''$ show no crossover, suppressing Maxwell-like behavior and confirming that reversible bond dynamics control the characteristic relaxation time.}
    \label{fgr:kb_dependence}
\end{figure}

With $k_b^0 = 0$, we effectively deactivate bond breakage. As a result, both the storage ($G'$) and loss ($G''$) moduli remain nearly constant across all frequencies, with no crossover point (Fig.~\ref{fgr:kb_dependence}). This confirms that the binding--unbinding dynamics control the characteristic relaxation time, $\tau_R$, of our system, consistent with theoretical predictions~\cite{parada_2018_ideal}.

\subsection{Appendix 3}\label{Appendix3}
\begin{figure}[htbp!]
    \centering
   \includegraphics[width=0.6\columnwidth]{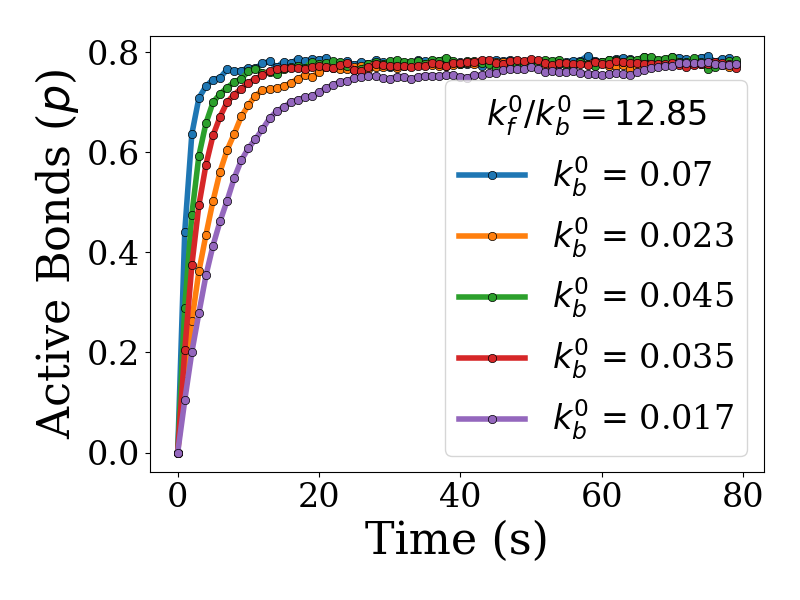}
    \caption{The equilibrium active bond fraction $p$ remains unchanged across all $k_b^0$ values}
    \label{fgr:varying_tau_comparison}
\end{figure}

To vary the bond relaxation time without changing the fraction of active bonds, $p$, we systematically varied $k_b^0$ and adjusted $k_f^0$ to maintain a constant ratio, $k_f^0/k_b^0 = 12.85$. This ensured that the equilibrium bond fraction remained constant across all values of $k_b^0$, while the time required to reach equilibrium changed. Specifically, faster bond-exchange dynamics led to more rapid equilibration to $p$ (Fig.~\ref{fgr:varying_tau_comparison}).

\bibliography{bibliography} 
\bibliographystyle{rsc} 

\end{document}